\documentclass[lettersize,onecolumn, journal]{IEEEtran}

\usepackage{amsmath,amssymb,amsfonts}
\usepackage{graphicx}
\usepackage{textcomp}
\usepackage[dvipsnames]{xcolor}
\usepackage{colortbl}
\usepackage{lipsum}

\usepackage{mathtools}  
\usepackage{amssymb}
\usepackage{tabulary}
\usepackage{booktabs}
\usepackage[inkscapearea=page]{svg}
\usepackage{auto-pst-pdf}

\usepackage{paralist}

\usepackage{bbding}
\usepackage{pifont}
\usepackage{wasysym}
\usepackage{amssymb}
\usepackage{diagbox}

\def\BibTeX{{\rm B\kern-.05em{\sc i\kern-.025em b}\kern-.08em
    T\kern-.1667em\lower.7ex\hbox{E}\kern-.125emX}}

\usepackage[top=0.75in, left=0.625in, bottom=1.in, right=0.625in]{geometry}

\usepackage[font=small]{caption}
\usepackage{subcaption}
\usepackage{booktabs}
\usepackage{bbm}
\usepackage{algorithm}
\usepackage{algcompatible}

\renewcommand{\COMMENT}[2][.1\linewidth]{%
  \leavevmode\hfill\makebox[#1][r]{ $\diamond$ \text{~#2}}}
\algnewcommand\algorithmicto{\textbf{to}}
\algnewcommand\RETURN{\State \textbf{return} }

\usepackage[numbers, compress]{natbib}

\usepackage{mathptmx}
\usepackage{amsmath}
\usepackage{amssymb}

\DeclareMathOperator*{\argmin}{arg\,min}

\newcommand\indicator{\mathbbm{1}}

\newcommand{\bb}[1]{{\color{blue}#1\color{blue}}}

\definecolor{mygreen}{RGB}{219, 48, 122}
\newcommand{\bred}[1]{{\color{BrickRed}\mathbf{#1}\color{black}}}
\newcommand{\bblue}[1]{{\color{Blue}\mathbf{#1}\color{black}}}
\newcommand{\bgreen}[1]{{\color{black}\mathbf{#1}\color{black}}}

\newcommand{\ie}{{\textit{i.e., }}}

\newcolumntype{P}[1]{>{\centering\arraybackslash}p{#1}}

\usepackage[colorlinks=true,allcolors=blue]{hyperref}

\usepackage{setspace}

\begin{document}

\title{
    \LARGE{Neural-Enhanced Rate Adaptation and Computation Distribution for Emerging mmWave Multi-User 3D Video Streaming Systems}
}
\author{
    \IEEEauthorblockN{Babak Badnava\IEEEauthorrefmark{1}, Jacob Chakareski\IEEEauthorrefmark{2}, Morteza Hashemi\IEEEauthorrefmark{1}} \\
    \IEEEauthorblockA{
    \IEEEauthorrefmark{1} Department of Electrical Engineering and Computer Science, University of Kansas \\
    \IEEEauthorrefmark{2} College of Computing, New Jersey Institute of Technology
    }
}
\maketitle
\thispagestyle{plain}
\pagestyle{plain}


\setlength{\textfloatsep}{0pt}
\setlength{\belowcaptionskip}{0pt}

\begin{abstract}
We investigate multitask edge-user communication-computation resource allocation for $360^\circ$ video streaming in an edge-computing enabled millimeter wave (mmWave) multi-user virtual reality system.
To balance the communication-computation trade-offs that arise herein, we formulate a video quality maximization problem that integrates interdependent multitask/multi-user action spaces and rebuffering time/quality variation constraints.
We formulate a deep reinforcement learning framework for \underline{m}ulti-\underline{t}ask 
\underline{r}ate adaptation and \underline{c}omputation distribution (MTRC) to solve the problem of interest.
Our solution does not rely on a priori knowledge about the environment and uses only prior video streaming statistics (e.g., throughput, decoding time, and transmission delay), and content information, to adjust the assigned video bitrates and computation distribution, as it observes the induced streaming performance online.
Moreover, to capture the task interdependence in the environment, we leverage neural network cascades to extend our MTRC method to two novel variants denoted as R1C2 and C1R2.
We train all three methods with real-world mmWave network traces and $360^\circ$ video datasets to evaluate their performance in terms of expected quality of experience (QoE), viewport peak signal-to-noise ratio (PSNR), rebuffering time, and quality variation.
We outperform state-of-the-art rate adaptation algorithms, with C1R2 showing best results and achieving $5.21-6.06$ dB PSNR gains, $2.18-2.70$x rebuffering time reduction, and $4.14-4.50$ dB quality variation reduction. 

\end{abstract}


\section{Introduction}\label{sec:Intro}
It is envisioned that next generation wireless networks (6G-and-Beyond) will enable an unprecedented proliferation of computationally-intensive and bandwidth hungry applications (e.g. Virtual/Augmented Reality (VR/AR), and online 3D gaming~\cite{SeaGate-2019-State}).
In particular, VR use-cases 
hold tremendous potential to advance our society and impact our daily life and the economy~\cite{ChakareskiKY:20,Chakareski-2023-Millimeter}.
Presently, VR applications are becoming increasingly popular in education, training, healthcare and gaming, 
reaching a global market size of \$32.64 billion in 2024~\cite{fortune-2024-vr}.

These emerging VR applications require streaming of high fidelity $360^\circ$ video content, which requires an ample amount of computational and communication resources near the edge of the network.
For instance, a minimum of $12$K high-quality spatial resolution and $100$ frame-per-second (FPS) temporal rate are recommended by MPEG for $360^\circ$ VR~\cite{Chakareski-2020-6DOF}.

\begin{figure}[t]
    \centering
    \includegraphics[width=.6\linewidth, trim={0cm 3cm 13.5cm 9cm},clip]{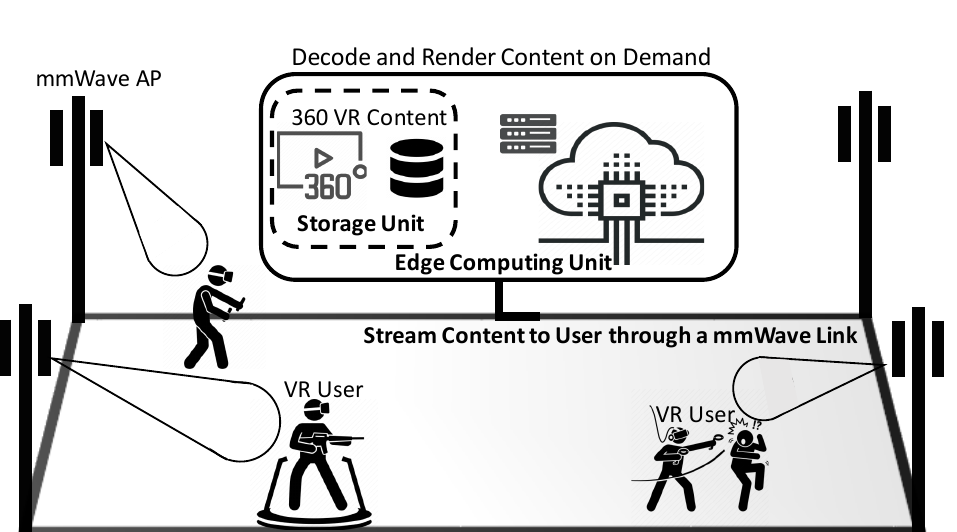}
    \caption{Edge-assisted VR system model: Multiple VR headsets connected to an edge computing unit (ECU) via a mmWave network.}
    \label{fig:system-model}
\end{figure}

Furthermore, in contrast to traditional 2D video streaming, $360^\circ$ video streaming  requires computational resources 
for encoding, decoding, spatial processing, stitching, and rendering~\cite{Chakareski-2020-6DOF}.
For instance, $360^\circ$ video decoding entails 
spatio-temporal transformations for spherical projection.
Viewport-adaptive streaming further increases the computational complexity by dynamically adjusting video segments based on the viewer's field of view (FoV). 
Moreover, $360^\circ$ videos typically have higher resolution and larger file sizes compared to 2D videos, leading to higher bandwidth requirements.
Hence, practical VR architectures face significant challenges including efficient resource management (computing and communication), coordination among distributed VR users, and providing guaranteed QoE.
Within this context, there are a few distinct sets of requirements that must be met:
(i) \emph{network-imposed} constraints that determine the available communication link rate,
(ii) \emph{computation-imposed} constraints that determine the available computational resources for processing $360^\circ$ videos,
(iii) \emph{video-imposed} constraints that vary depending on the \emph{spatio-temporal characteristics} of $360^\circ$ videos,
(iv) \emph{user-imposed} constraints that determine the user's preference in terms of QoE.
To address these challenges, we pursue a novel multitask edge-assisted video quality maximization framework for multi-user $360^\circ$ video streaming in mmWave networks, as depicted in Fig. \ref{fig:system-model}.
Here, $360^\circ$ videos can be decoded and/or rendered by an edge computing unit (ECU), located on the premise of the VR arena.
On one hand, the ECU has more computational resources, thus can process (\ie decode/render) the $360^\circ$ videos faster, which leads to a lower computation latency and a higher QoE for users.
On the other hand, decoding and rendering at ECU introduces a higher bandwidth requirement for each user since processed videos have much larger sizes, which leads to higher communication latency, thus degrading the QoE.
Moreover, the VR system is a stochastic system due to time-varying network conditions and varying spatio-temporal characteristics of $360^\circ$ videos.
Our proposed multitask decision-making framework considers the interplay between the communication and computation requirements of a mobile $360^\circ$ video streaming application.
Leveraging a state-of-the-art deep reinforcement learning (DRL) method~\cite{cobbe-2020-ppg} and multitask neural network cascades~\cite{Dai-2016-Instance}, we present three DRL agent architectures, called MTRC, C1R2, and R1C2. Our proposed solutions solve a constrained multitask video quality maximization problem that integrates user performance requirements (\ie rebuffering time and viewport quality variation). These agents learn the optimal rate allocation and computation distribution (\ie ECU or headset) in the VR arena by considering the playback statistics (\ie past throughput, decoding time, transmission time, etc.) and content information.

The main contributions of our paper are:
\begin{enumerate}
    \item \emph{Constrained multitask video quality maximization problem.}
    We introduce a multi-user edge-assisted VR streaming schema, where an ECU serves VR users with $360^\circ$ videos.
    Then, we formulate a constrained multitask video quality maximization problem to find the best rate adaptation and computation distribution policy w.r.t. users' performance requirements, network condition, and spatio-temporal characteristics of $360^\circ$ videos.
    \item \emph{Multitask rate adaptation and computation distribution.}
    We develop a learning-based multitask rate adaptation and computation distribution framework, in which we introduce three DRL agent architectures (\ie MTRC, R1C2, and C1R2) to find the optimal rate adaptation and computation distribution policy that maximizes streamed video quality while meeting users' performance requirements.
    \item \emph{Multi-user $360^\circ$ video streaming simulator.}
    We develop a trace-driven gym-like~\cite{brockman2016openai} $360^\circ$ VR streaming simulator using real-world datasets of $360^\circ$ video rate-distortion characteristics and user head movement navigation information, as well as mmWave network traces.
    \item \emph{Extensive numerical analysis.}
    Leveraging the developed simulator, we perform an extensive numerical analysis of our proposed methods in various system conditions.
    We show that our proposed solutions learn to balance the existing trade-offs in the system and outperform the state-of-the-art rate adaptation algorithm. 
    Specifically, our proposed methods demonstrate $5.21-6.06$ dB PSNR gains, $2.18-2.70$x rebuffering time reduction, and $4.14-4.50$ dB quality variation reduction. 
    
    
\end{enumerate}
The rest of our paper is organized as follows. 
In Section \ref{sec:related}, we review related work and highlight our contributions.
In Section \ref{sec:system}, we present the multi-user edge-assisted VR system model, followed by the constrained multitask video quality maximization problem in Section \ref{sec:problem}.
In Section \ref{sec:solution}, we present our learning-based multitask rate adaptation and computation distribution framework.
In Section \ref{sec:evaluation}, we provide simulation results and analysis of the performance of the proposed architectures. Section \ref{sec:conclusion} concludes the paper.


\section{Related Work}\label{sec:related}

{\bf MEC-Assisted VR Systems.} Integrating multi-access edge computing (MEC) in VR systems has been explored to provide additional: (i) computational power for decoding, rendering, and stitching of $360^\circ$ videos \cite{Chakareski-2020-6DOF, Hsu-2020-MEC} and (ii) storage space for caching $360^\circ$ video content~\cite{Dai-2020-View, Maniotis-2021-Tile,Dang-2019-Joint}.
Moreover, MEC node placement \cite{Zhang-2022-UAV,Liu-2023-On}, architectures for on demand/real-time streaming \cite{Guo-2021-Design, Aung-2024-Edge,Han-2019-Real,Zhang-2023-RealVR}, communication resource management \cite{Yang-2018-Communication, Lin-2021-Resource, Guo-2020-AnAW, Chen-2019-Data}, and user scheduling \cite{Huang-2018-MAC} have been explored. Finally, mobile edge computing integrated with high-bandwidth mmWave links and multi-connectivity \cite{Chakareski-2024-Live, Ge-2017-Multipath, Liu-2019-MEC, Gupta-2023-mmWave, Ren-2019-Edge, Chakareski-2023-Millimeter} has been studied to enable multi-Gbps data rates needed for lifelike VR immersion. 
We advance these studies by formulating a dynamic DRL decision-making framework for adaptive rate and computational task allocation in a mmWave VR network system. 
Our framework establishes a \emph{user-edge computing continuum}, which dynamically makes decoding and rendering decisions for each video segment under time-varying mmWave channel conditions. This, in turn, enhances the flexibility of the system to further improve users'
 QoE by taking into account the spatio-temporal characteristics of $360^\circ$ videos.

\textbf{Rate Adaptation for VR Systems.} 
The authors in \cite{Hou-2021-Predictive} proposed a FoV prediction algorithm, where the predicted view is encoded at relatively high quality and transmitted in advance to alleviate the network-imposed constraints by reducing the bandwidth requirements, leading to smaller delays. Other FoV prediction methods have been explored in 
\cite{Chen-2021-Sparkle,Chakareski-2023-User,Fan-2020-Optimizing} with the same objective.
Moreover, 
rate adaptation methods have been integrated with viewport prediction to further save bandwidth and enhance the QoE~\cite{Kumar-2023-Seer, Petrangeli-2017-Improving, Kan-2020-RAPT360, Cagri-2017-Viewport}. We note that 
we assume the availability (or prediction) of viewport information in our approach.
Finally, there are several works that studied joint computation and communication design~\cite{Cheng-2022-Design, wang2023bones, Hieu-2023-When, Liu-2023-Graph, Li-2018-MUVR, Chen-2022-Wireless}. However, they differ from our framework either in terms of the objective (end-to-end latency \cite{Cheng-2022-Design, Hieu-2023-When, Chen-2022-Wireless}, computational load efficiency \cite{Li-2018-MUVR}), or the decision variables they consider (network bandwidth \cite{wang2023bones}, optimal computing node \cite{Liu-2023-Graph}). 
While these studies focus on alleviating the communication imposed constraints, our work advances these studies 
by 
employing the additional computational resources provided by edge computing to  balance the communication and computation trade-offs arising in a networked VR system. 

Our paper extends our preliminary results \cite{Babak-2024-Multi}, by proposing two new multitask RL agents (R1C2 and C1R2) based on neural network cascades, which effectively capture the interdependence between the decision variables in the formulated optimization problem. We have also significantly extended our numerical evaluations in various directions.
We first demonstrate further improvement by neural network cascades solutions (i.e., R1C2 and C1R2), in terms of video quality, rebuffering time, and quality variation perceived by VR users.
Moreover, we investigate the trade-offs between these QoE metrics.
Finally, we provide more detailed performance evaluation of our proposed methods under various network conditions (low throughput, medium throughput, high throughput).


\section{Edge-Assisted VR System Model}\label{sec:system}
As depicted in Fig. \ref{fig:system-model}, we consider a multi-user $360^\circ$ VR video streaming application with $N$ users  
that are equipped with VR headsets and  connected to an ECU through  mmWave wireless links.
All VR headsets and the ECU are equipped with computational resources (\ie CPU and GPU) to decode/render multi-layer $360^\circ$ videos, as described next. 

\begin{figure}[t]
    \centering
    \includegraphics[width=.9\linewidth, trim={4.3cm 5.3cm 4.3cm 3.8cm},clip]{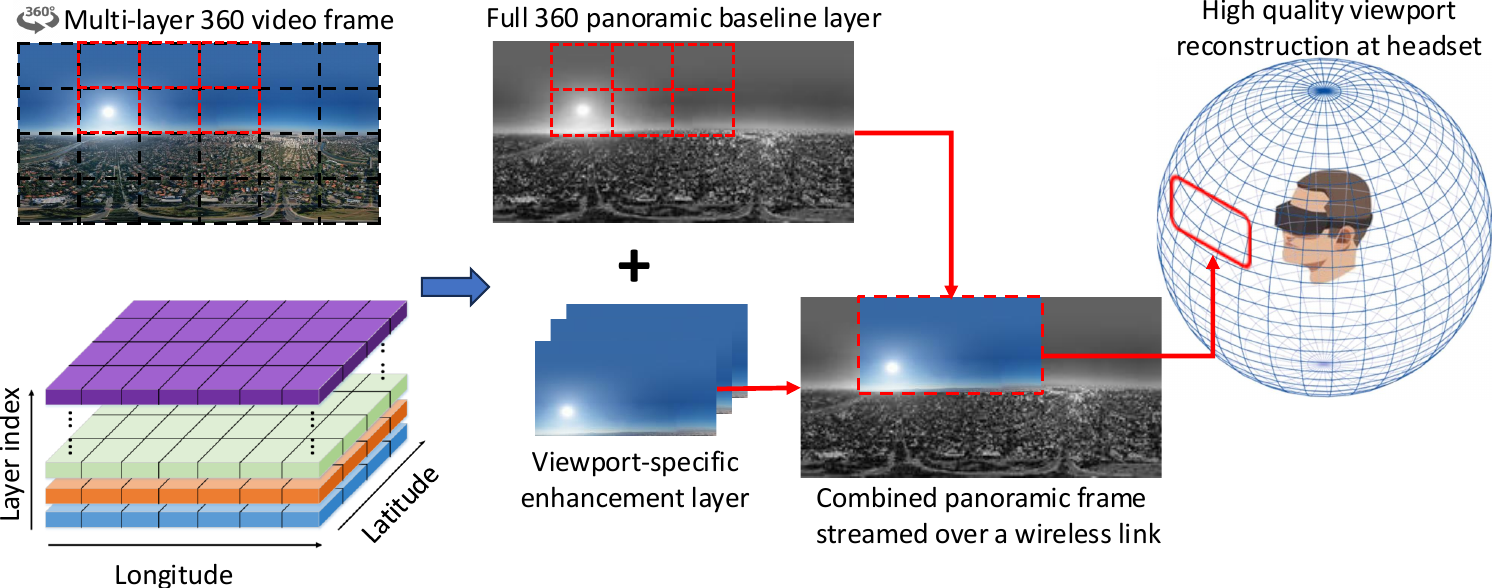}
    \caption{Multi-layer $360^\circ$ video model (lower left corner): Viewport-specific enhancement layers combined with a wide $360^\circ$ panorama baseline layer are transmitted to a VR headset via a mmWave wireless link. As the number of added enhancement layers increases, the video bitrate and the video quality delivered to the VR user increase.}
    \label{fig:video-model}
\end{figure}
\textbf{Multi-layer $360^\circ$ Video Model:}
We consider a set of $360^\circ$ videos stored on ECU and leverage the scalable multi-layer $360^\circ$ video viewpoint tiling design~\cite{Chakareski-2020-6DOF}.
As depicted in Fig. \ref{fig:video-model}, each panoramic $360^\circ$ video frame is partitioned into $\mathcal{L}$ tiles arranged in a $L_H \times L_V$ grid.
A block of consecutive video frames, compressed together with no reference to others, creates a group of pictures (GoP) or video chunk.
Each video is divided into $M$ GoP with fixed time duration of $\Delta t$.
The ECU constructs $L$ layers of increased immersion fidelity for each tile in a GoP.
The first layer is called the base layer, and the remaining layers are denoted as enhancement layers. 
Each enhancement layer incrementally increases the transmission video bitrate and subsequently the video quality delivered to the user.
For each GoP, ECU constructs a baseline representation of the entire $360^\circ$ panorama by using the base layer for all tiles.
Then, ECU constructs an enhancement representation by integrating the enhancement layers for the tiles within the user's current viewport.
Finally, these two representations are combined with each other and streamed to the VR headset over a mmWave wireless link.
We denote $e^m_n \in \{0, 1\}^{L}$ as the number of enhancement layers included in the $m^{th}$ GoP requested by the $n^{th}$ user.
Then, the size of the $m^{th}$ compressed GoP, denoted by $d(e^m_n)$, is determined by summing over all tiles' bitrates in the GoP.
We assume a positive compression reduction factor of $\beta < 1 $, which leads to $d(e^m_n)/\beta$ for the size of the $m^{th}$ decoded GoP.
After decoding, GoPs need to be rendered as well, which leads to an increase in size by a factor of $\alpha \geq 2$.
Hence, the size of the $m^{th}$ GoP after decoding and rendering would be $\alpha d(e^m_n)/\beta$.

\subsection{User Headset Model}
The VR headset is connected to the ECU via a mmWave wireless link for video streaming.
The ECU decides on sending raw (\ie compressed), decoded, or decoded and rendered video to each headset.
The VR headsets are also equipped with a CPU and GPU to process the videos in case the ECU decides to send the raw video to the user. We assume that the $n^{th}$ VR headset provides maximum decoding and rendering speeds of $\bar{Z}^{dec.}_{n}$ and $\bar{Z}^{rend.}_{n}$, respectively. 
The VR headset buffers the rendered video in a buffer with fixed time duration length, as described next. 

\begin{figure}[t]
    \centering
    \includegraphics[width=.6\linewidth, trim={0cm 0.4cm 0cm 0cm},clip]{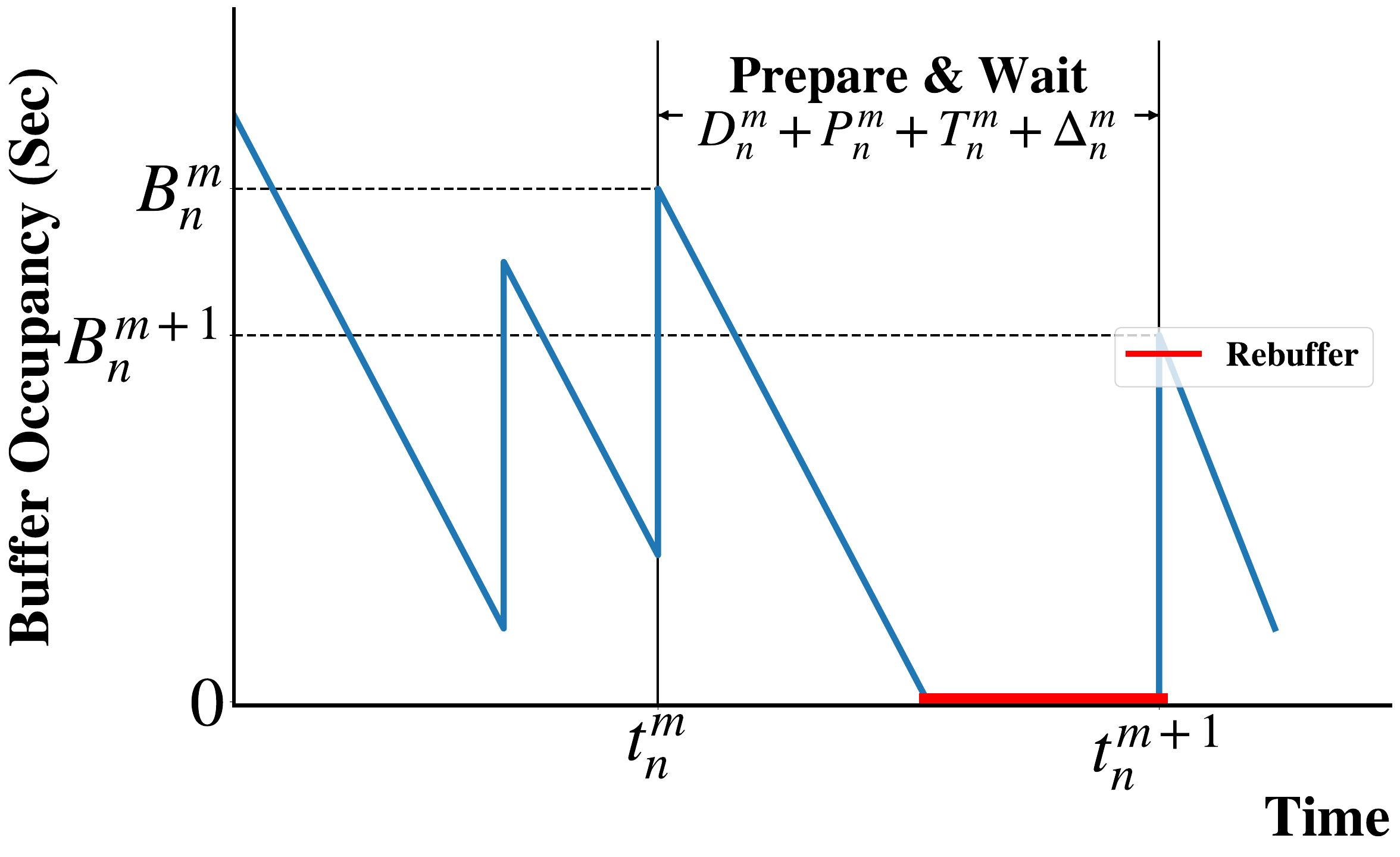}
    \vspace{-.25cm}
    \caption{VR headset playback buffer dynamics.}
    \label{fig:buffer-dynamics}
\end{figure}

\textbf{Playback Buffer Dynamics:}
Fig. \ref{fig:buffer-dynamics} illustrates the buffer dynamics of the $360^\circ$ video streaming application.
Let $B_n(t^m_n)$ denote the buffer occupancy of user $n$ at time $t^m_n$, when the $m^{th}$ GoP is requested by the $n^{th}$ user. For the sake of exposition, we use $B^m_n$ to refer to $B_n(t^m_n)$. 
Then, the buffer occupancy is modeled as:
\begin{equation}
\small
\begin{aligned}
    B^{m+1}_n = 
    \left(
        \left(
        B^m_n  - D^m_n - P^m_n - T^m_n
        \right)_+
        + \Delta t - \Delta^m_n
    \right)_+.
\end{aligned}
\end{equation}
Here  $\left(x\right)_+ = \max \left\lbrace 0, x \right\rbrace$ ensures the non-negativity of the buffer occupancy.  
$D^m_n$, $P^m_n$, and $T^m_n$ denote the decoding, rendering, and transmission times of the GoP. We note that the transmission time $T^m_n$ is a function of the mmWave network condition and the GoP's size, which varies depending on whether the GoP is in its raw, decoded, or rendered state. Furthermore, $\Delta t$ is the fixed GoP duration, which will be added to the buffer occupancy level after GoP $m$ is completely received.
$\Delta^m_n$ denotes the waiting time after GoP $m$ is completely received. The reason for incorporating the waiting time $\Delta^m_n$ is due to the fact that the playback buffer could be full, and thus the headset may need to wait until the playback buffer has enough space to accommodate the next GoP, which leads a waiting time of:
\begin{equation}
\small
\begin{aligned}
        \Delta^m_n = \left( 
            \left( B^m_n - P^m_n - D^m_n - T^m_n \right)_+ 
            + \Delta t - B^{max}_n
        \right)_+.
\end{aligned}
\end{equation}
Here, $B^{max}_n$ is the maximum buffer occupancy level of $n^{th}$ user.
Therefore, the request time for the next GoP is obtained as: 
\begin{equation}
\small
\begin{aligned}
    t^{m+1}_n = t^m_n + D^m_n + P^m_n + T^m_n + \Delta^m_n.
\end{aligned}
\end{equation}

When the preparation phase takes more time than the amount of video left in the buffer (\ie $B^m_n < P^m_n + D^m_n + T^m_n$), the buffer becomes empty, leading to a \textit{rebuffering event}, as shown in Fig. \ref{fig:buffer-dynamics}. The rebuffering time for the $m^{th}$ GoP of user $n$ is:
\begin{equation}
\small
S_n^m = \left( D^m_n + P^m_n + T^m_n - B^m_n \right)_+.
\end{equation}
In Section \ref{sec:problem}, we will use $S_n^m$ to formulate the aggregate rebuffering time for the $360^\circ$ video streamed to user $n$.

\textbf{VR Headset Decoding and Rendering Model:}
We incorporate a decoding and rendering model developed by \cite{Chakareski-2023-Millimeter} to compute the decoding and rendering time for each GoP.
In this model, the decoding time of the $m^{th}$ GoP for the $n^{th}$ user is assessed as $\Tilde{si}(e^m_n)/ \bar{Z}^{dec.}_n$, where $\Tilde{si}(.)$ returns the decoding computational complexity of a GoP, in bits, which is the induced data rate associated with the current viewport and the quality of the streamed video (\ie $\Tilde{si}(e^m_n) = d(e^m_n)$).
Similarly, the rendering time is modeled as $si(e^m_n)/ \bar{Z}^{rend.}_n$, where $si(.)$ returns the rendering computational complexity of a GoP, in bits, which is the induced data rate after decoding of the GoP (\ie $si(e^m_n) = d(e^m_n) / \beta$).
Note that we assume that the viewport information is available on the headset.


\subsection{Edge Computing Unit Model}
The ECU is equipped with a GPU, which provides  maximum decoding and rendering speeds of 
$Z^{dec.}_{ECU}$ and $Z^{rend.}_{ECU}$, respectively. 
The computational resources of the ECU (\ie decoding and/or rendering resources) are shared among users to provide extra computational power for users to decode/render their videos.

\textbf{Video Decoding and Rendering Model:}
The decoding starts immediately after receiving the request for the next GoP if the decision-maker decides to decode the GoP on the ECU. This leads to
$
            \Tilde{si}(e^m_n) /
            \psi^m_n
$
seconds of decoding time.
Here, $\psi^m_n$ denotes the amount of decoding resources allocated to the $n^{th}$ user for decoding the $m^{th}$ GoP.
Similarly, the rendering time of the ECU is modeled as $si(e^m_n) / \theta^m_n$, where $\theta^m_n$ denotes the amount of rendering resources allocated to the $n^{th}$ user for the $m^{th}$ GoP.
Note that the total amount of computational resources allocated to users cannot exceed the maximum available resources, which means that $\sum_{n=1}^N \psi^m_n \leq Z^{dec.}_{ECU}$ and $\sum_{n=1}^N \theta^m_n \leq Z^{rend.}_{ECU}$ must be satisfied for each GoP.

\textbf{Communication Model:}
The ECU transmits GoPs through a mmWave wireless network.
The expected transmission rate for a GoP is modeled as
$
C^m_n = \frac{1}{t_e - t_s} \int_{t_s}^{t_e} C^s_n ds,
$
where $t_s$ and $t_e$ are the transmission start and end times, respectively, and $C^s_n$ is the throughput provided by the mmWave channel for the $n^{th}$ user.
Hence, the transmission time for a compressed GoP follows
$
d(e^m_n) / C^m_n.
$
Similarly, we can express the transmission time for decoded and rendered GoPs, but are omitted here for brevity.

\section{Constrained Video Quality Optimization}\label{sec:problem}
Here, we formulate the multitask constrained video quality maximization problem for $360^\circ$ video delivery. We first examine the factors that impact the QoE in such a setting.

\textbf{Quality of Experience (QoE):}
One of the key goals for VR systems is to improve the QoE that can lead to increased long-term user engagement.
While users may differ in their preference, there are three key contributing factors to the QoE: 
\begin{enumerate}
    \item \textit{Average Video Quality (AVQ) :} The average per-GoP video quality for tiles in user's FoV: $Q(\mathbf{e_n})=\frac{1}{M} \sum_{m=1}^{M} q(e^m_n)$.
    \item \textit{Average Quality Variation (QV):} The average quality variation in the user's FoV from one GoP to another: $V(\mathbf{e_n}) = \frac{1}{M-1} \sum_{m=1}^{M-1} \left|q(e^{m+1}_n) - q(e^m_n)\right|$.
    \item \textit{Rebuffering Time (RT):} Rebuffering occurs if the delivery time for a GoP is higher than the buffer occupancy level when that GoP was requested. Thus, the aggregate rebuffering time for the entire $360^\circ$ video streamed to user $n$ is given by: $S(\mathbf{e_n}, \mathbf{\phi_n}) = \sum_{m=1}^{M} S_n^m$.
    
\end{enumerate}
Here, $\mathbf{\phi_n} = \{\phi^1_n, ..., \phi^M_n\}$ and $\mathbf{e_n} = \{e^1_n, ..., e^M_n\}$, respectively, denote the computation distribution and rate adaptation decisions of the $n^{th}$ user for a video with $M$ GoPs.
While there are various choices for the function $q(.)$, 
we use PSNR for the viewer's FoV, which can be calculated using the video distortion as $q(e^m_n) = 10 \log_{10}(255^2 / MSE^m_n)$, where $MSE^m_n$ is the distortion of the $m^{th}$ GoP. The distortion has an inverse relation with the video bitrate, which is determined by the number of enhancement layers streamed to the user~\cite{Chakareski-2021-Full, Yu-2015-Framework, Mao-2017-Neural}.
Several prior studies \cite{wang2023bones, Mao-2017-Neural, Spiteri-2020-BOLA, Yin-2015-Control} define the QoE of the user by a weighted sum of the aforementioned components:
\begin{equation}
\label{eq:qoe}
\begin{aligned}
    \hat{QoE}_n^M = Q(\mathbf{e_n})   -  \mu_0 S(\mathbf{e_n}, \mathbf{\phi_n}) - \mu_1 V(\mathbf{e_n}),
\end{aligned}
\end{equation}
where $\mu_0$ and $\mu_1$ are non-negative weighting parameters corresponding to user sensitivity to rebuffering time and quality variation, respectively.
While this QoE metric allows us to model varying user preferences \cite{Mao-2017-Neural, Yin-2015-Control}, 
it is challenging to determine the exact values of the coefficients $\mu_0$ and $\mu_1$ for various users' requirements.

\textbf{Constrained Video Quality Optimization Problem:}
To formulate our problem of multitask video quality maximization, we define two sets of communication and computation decision variables.  
In particular, $\phi^m_n \in \left\{ 0, 1 \right\}^{3}$ is a binary vector of size three with one active element (\ie a one-hot vector), which determines where the decoding and rendering take place for each GoP and user.
There are three distinct states for $\phi^m_n$:
 \begin{equation}
 \label{eq:phi-states}
 \small
  \phi^m_n: 
  \begin{cases}
        \phi^{m, 0}_n = 1  \quad \Rightarrow \; \text{Decode \& Render on ECU,} \\
        \phi^{m, 1}_n = 1 \quad \Rightarrow \; \text{Decode on ECU \& Render on headset,}\\
        \phi^{m, 2}_n = 1 \quad \Rightarrow \; \text{Decode \& Render on headset.}
        \end{cases}
 \end{equation}
In addition to computation location, we consider the rate allocation decision variable $e^m_n \in \{0, 1\}^{L}$ that determines how many enhancement layers should be streamed to the $n^{th}$ user in the VR arena. 
In addition to these decision variables, ECU computation resources (\ie decoding resources $\psi_n^m$  and rendering resources $\theta_n^m$) should be allocated to the users,
which we assume that they have been allocated proportional to user's requirements. 
Therefore, we formulate the optimization problem outlined in Eq. \eqref{eq:qoe-maximization} for each VR user:

\begin{figure}[ht]
\hrulefill
\begin{subequations}\label{eq:qoe-maximization}
\begin{alignat}{5}
    \max_{\mathbf{\phi_n, e_n}} \quad\;  Q(\mathbf{e_n}) & \tag{\ref{eq:qoe-maximization}} \\
    \text{s.t.}
    \quad  S(\mathbf{e_n}, \mathbf{\phi_n}) & \leq \mathcal{H}_0 \label{eq:rt-const} \\
    \quad  V(\mathbf{e_n})  & \leq \mathcal{H}_1 \label{eq:qv-const} \\
    \quad B^{m+1}_n &= 
    \left(
        \left(
        B^m_n - P^m_n - D^m_n - T^m_n
        \right)_+
        + \Delta t - \Delta^m_n
    \right)_+ \; \forall m \label{eq:buffert-b}\\
    \quad   t^{m+1}_n &= t^m_n \nonumber\\
    \quad &+ \phi^{m,1}_n \left[ 
    \frac{
            \Tilde{si}(e^m_n)
        }{
            \psi^m_n
        } + \frac{
            si(e^m_n)
        }{
            \theta^m_n
        } + \frac{
             \alpha  d(e^m_n)/\beta 
        }{
            C^m_n
        } \right]
    \nonumber \\ &+ \phi^{m,2}_n \left[
        \frac{
            \Tilde{si}(e^m_n)
        }{
            \psi^m_n
        } + 
        \frac{
            si(e^m_n)
        }{
            Z^{rend.}_{n}
        } + 
        \frac{
            d(e^m_n)/\beta
        }{
            C^m_n
        } \right] \nonumber \\
    \quad &+ \phi^{m,3}_n \left[
        \frac{
            \Tilde{si}(e^m_n)
        }{
            Z^{dec.}_{n}
        }
    +   \frac{
            si(e^m_n)
        }{
            Z^{rend.}_{n}
        } + 
        \frac{
            d(e^m_n)
        }{
            C^m_n
        }
    \right] 
    + \Delta^m_n  \; \forall m  \label{eq:buffert-t} \\
    \quad \sum_{n=1}^N  \psi^m_n &\leq Z^{dec.}_{ECU} \quad \forall m \quad, 
    \quad \sum_{n=1}^N \theta^m_n \leq Z^{rend.}_{ECU} \quad \forall m \label{eq:comp-const}
\end{alignat}
\end{subequations}
\hrulefill
\end{figure}
\noindent 
Here,
Eq. \eqref{eq:rt-const} makes sure that the rebuffering time is bounded by $\mathcal{H}_0$, and
Eq. \eqref{eq:qv-const} guarantees that the quality variation is less than or equal to $\mathcal{H}_1$.
Furthermore, Eq. \eqref{eq:buffert-b} and Eq. \eqref{eq:buffert-t} define the user's playback buffer dynamics,
and Eq. \eqref{eq:comp-const} is related to the computation constraints on the ECU.

The optimization problem in Eq. \eqref{eq:qoe-maximization} aims to maximize the quality of video delivery while limiting rebuffering time and quality variations. To solve this problem, there are several key challenges, including: {\emph{(i) Multitask Interdependent Action Space:}}
    The action space contains two discrete sub-actions (\ie rate adaptation and computation distribution).
    The combination of these two creates a combinatorial action space, which makes it challenging to find the optimal action for each GoP.
    Furthermore, sub-actions are interdependent, such that the computation distribution action $\phi^m_n$ puts a limit on the other actions. For example, decoding and rendering a GoP on a headset limits the maximum achievable rate on the headset due to limited computational resources. {\emph{(ii) Time-Varying System Characteristics:}} 
    The VR environment is dynamic due to varying $360^\circ$ video characteristics and time-varying network conditions.
    The amount of computational resources required for decoding and rendering a $360^\circ$ video varies from one video to another depending on the quality of the video, encoding algorithm, etc. \cite{Chakareski-2020-6DOF}. 
    Moreover, VR users are connected to the ECU via a mmWave wireless network, which enables multi-Gbps transmission rates. However, this comes with the challenge of highly dynamic channels due to blockage, mobility, and interference \cite{Ghazikor-2024-Interference}. 
    {\emph{(iii) Multi-User Shared Environment:}}
    There are several users in the VR arena with different communication, computation, and performance requirements.
    As denoted in Eq. \eqref{eq:comp-const}, users share the ECU computational resources with each other. This means  a change in one of the users' requirements may lead to a performance degradation for other users. The ECU has limited computational resources that must be efficiently allocated to a subset of users for ECU-assisted computations such that the overall system performance is improved.  

Our goal is to develop a dynamic decision-making algorithm that addresses the challenges mentioned above.
Learning-based decision-making methods, DRL in particular, have shown promising performance in solving decision-making tasks in various applications~\cite{cobbe-2020-ppg,Mao-2017-Neural,schulman-2017-proximal,Deheng-2020-Mastering,Badnava-2023-Energy}.
The success of the learning-based methods stems from the fact that they do not rely on predefined models of the environment, but instead learn the optimal policy through repeated interactions with the environment and collecting rewards that are commensurate with the quality of the actions.  
DRL methods are generally capable of capturing uncertainty in the system due to varying $360^\circ$ video characteristics and time-varying network conditions. 
However, 
DRL methods, in general, are not well-suited for optimization problems with inequality constraints. Furthermore, 
multiple interdependent decision variables and shared computational resources in a multi-user environment introduce more challenges for DRL methods that need to be addressed.
Next, we propose a multitask decision-making framework to address these challenges and enhance QoE for all users in the VR arena.



\section{Multitask Resource Allocation Framework}\label{sec:solution}
To tackle the above challenges, we develop a multitask resource allocation solution using deep reinforcement learning (DRL) agent, which is realized using three different neural network architectures. 
Next, we describe the decision-making flow for each GoP, and then present the details of the MTRC agent (\ie state and action description, architecture of the neural networks, and training methodology).

\textbf{Decision-Making Flow and State:}
At each time step and after receiving the request for the $m^{th}$ GoP from all users,
the DRL agent observes the state $s^m$, which includes \emph{playback statistics and video information}.
The video playback statistics include six critical metrics for each user to fully describe the $360^\circ$ video playback status.
These metrics are past throughput $C^{m-1}_n$, past decoding time $D^{m-1}_n$, past transmission time $T^{m-1}_n$, past rendering time $P^{m-1}_n$, past allocated rate, and current buffer level $B^m_n$.
We take the future GoPs' size and number of remaining GoPs for each user as video information.
This will help the DRL agent to distinguish between videos with different spatio-temporal characteristics.
All of these metrics are collected for all users and stacked together to represent the state information of the system. 

\textbf{Action:}
Once the state $s^m$ is observed, the agent chooses an action $a^m$ to decode, render, and transmit GoPs.
In this case, the agent takes a joint action $a^m=(e^m, \phi^m)$.
The rate allocation action $e^m \in \{0, 1\}^{N\times L}$ determines how many enhancement layers will be streamed to each user, and the computation distribution action $\phi^{m} \in \{0,1\}^{N\times 3}$ determines where each user decodes and/or renders their $m^{th}$ GoP according to Eq. \eqref{eq:phi-states}.

\textbf{Reward:}
After taking the action,  the state of the environment changes, and the agent receives a reward vector $r^m$ that gives the reward for each user.
The goal of our agent is to maximize the expected cumulative discounted reward for all users.
To define the reward corresponding to the constrained optimization problem in Eq.~\eqref{eq:qoe-maximization},
we modify the optimization problem to capture the constraint violation and penalize the DRL agent for not meeting the constraints.
Leveraging the duality principle, we write the Lagrangian dual problem of Eq. \eqref{eq:qoe-maximization}:
\begin{equation}
\label{eq:qoe-lagrange}
\small
\begin{aligned}
    \min_{\mu_0, \mu_1} \; \max_{\mathbf{\phi_n, e_n}} & \;\;   \underbrace{Q(\mathbf{e_n})
    + \overbrace{\mu_0 \left( \mathcal{H}_0 - S(\mathbf{e_n}, \mathbf{\phi_n}) \right)}^{\mathcal{L}^{\mu_0}} 
    + \overbrace{\mu_1 \left( \mathcal{H}_1 - V(\mathbf{e_n}) \right)}^{\mathcal{L}^{\mu_1}}
    }_{:= QoE_n^M}\\
    \text{s.t.} & \;\; \text{Eqs.} \;\; \eqref{eq:buffert-b}, \eqref{eq:buffert-t}, \eqref{eq:comp-const},
\end{aligned}
\end{equation}
which is equal to:
\begin{equation}
\label{eq:qoe-lagrange-mu}
\small
\begin{aligned}
    \min_{\mu_0, \mu_1} & \quad\;   Q(\mathbf{e_n^\ast})   
    + \mu_0 \left( \mathcal{H}_0 - S(\mathbf{e_n^\ast}, \mathbf{\phi_n^\ast}) \right) 
    + \mu_1 \left( \mathcal{H}_1 - V(\mathbf{e_n^\ast}) \right)\\
    \text{s.t.} & \quad \text{Eqs.} \;\; \eqref{eq:buffert-b}, \eqref{eq:buffert-t}, \eqref{eq:comp-const}.
\end{aligned}
\end{equation}
Here, $\mathbf{e_n^\ast}$ and $\mathbf{\phi_n^\ast}$ are optimal rate allocation and computation distribution actions for the user $n$.
Thus, the optimal coefficients for rebuffering time and quality variation, which correspond to the Lagrangian multipliers, can be obtained by solving Eq. \eqref{eq:qoe-lagrange}:
\begin{equation}
\label{eq:mu-solutions}
\small
\begin{aligned}
    \mu_0^\ast &= \argmin_{\mu_0} \;\; \mu_0 \left( \mathcal{H}_0 - S(\mathbf{e_n^\ast}, \mathbf{\phi_n^\ast}) \right), \\
    \mu_1^\ast &= \argmin_{\mu_1} \;\; \mu_1 \left( \mathcal{H}_1 - V(\mathbf{e_n^\ast}) \right).
\end{aligned}
\end{equation}
Here, by minimizing the loss functions presented in Eq. \eqref{eq:mu-solutions}, $\mu_0$ and $\mu_1$ are updated to increase (or decrease) if the rebuffering time (or quality variation) is higher (or lower) than the target rebuffering time $\mathcal{H}_0$ (or video quality variation $\mathcal{H}_1$).
This modification in the optimization problem effectively handles the reward magnitude changes over time during training. 
Hence, one needs to set only the target rebuffering time $\mathcal{H}_0$ and the target quality variation $\mathcal{H}_1$ for each user and video. Then, the coefficients $\mu_0$ and $\mu_1$ are dynamically adjusted over time to meet the target constraints.

The objective function in Eq. \eqref{eq:qoe-lagrange} accounts for both rebuffering time and quality variation according to users' preference. Then, we employ the changes in users' perceived QoE at each step
as the reward term, which leads us to:
\begin{equation}
r^{m+1}_n = QoE^{m+1}_n - QoE^m_n.
\end{equation}
This reward captures the changes in the perceived QoE as a result of the last action performed by the rate adaptation and computation distribution agent, thereby enabling the agent to learn the actions that lead to improvement in QoE.

\begin{figure}
    \centering
    \includegraphics[width=.65\linewidth, trim={0cm 1.8cm 5.4cm 0cm},clip]{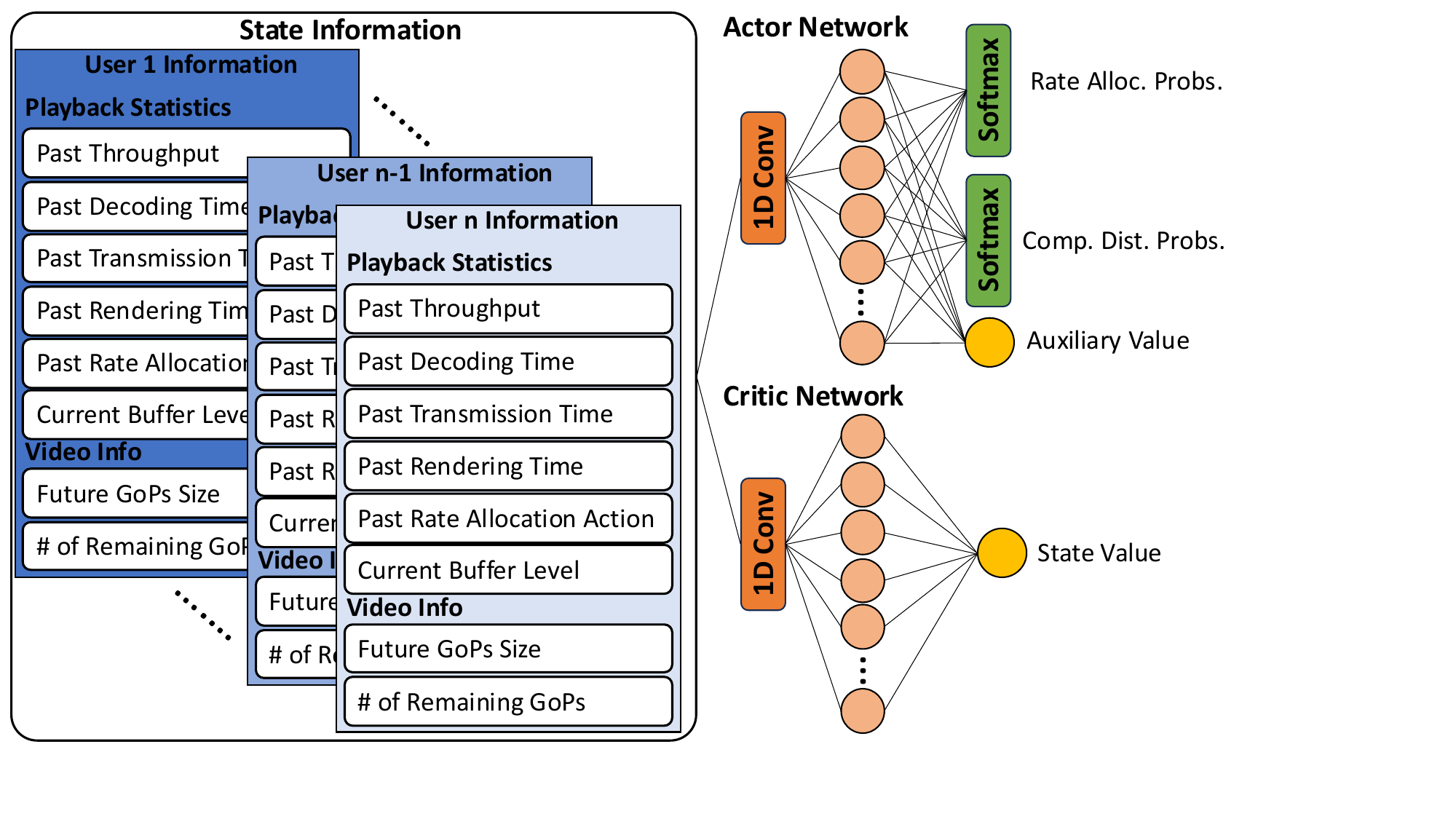}
    \caption{Our MTRC architecture comprises actor and critic networks. Based on the observed state information for each user, the actor network makes a joint rate adaptation and computation distribution decision, and the critic network estimates the state values.}
    \label{fig:deep-vr-arch}
\end{figure}

\begin{figure*}
    \centering
    \begin{subfigure}[t]{0.32\linewidth}
        \includegraphics[width=.95\linewidth]{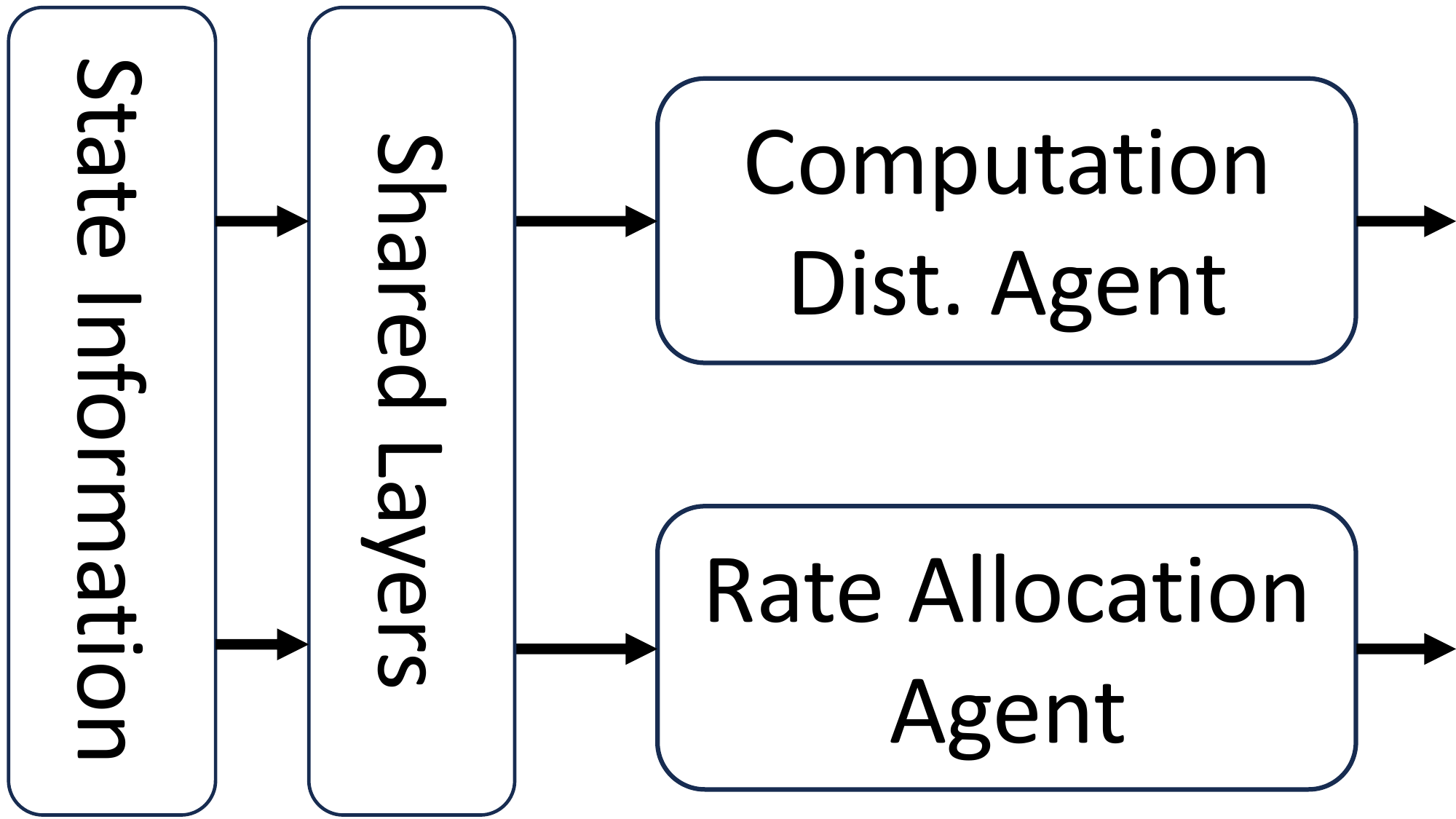}
        \caption{MTRC Agent Architecture}
        \label{fig:mtrc-arch}    
    \end{subfigure}
    \begin{subfigure}[t]{0.32\linewidth}
        \includegraphics[width=.95\linewidth]{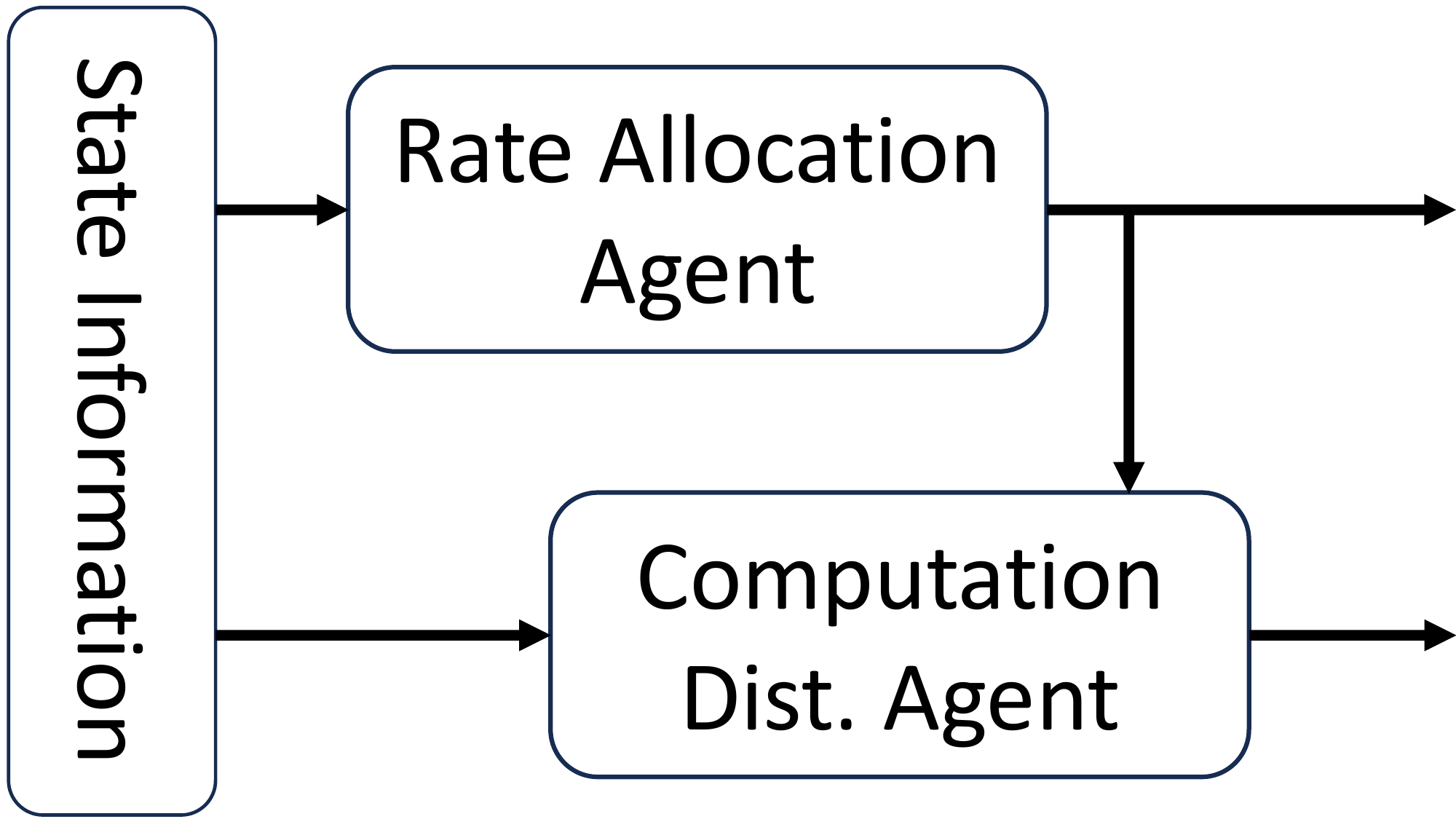}
        \caption{R1C2 Agent Architecture}
        \label{fig:r1c2-arch}
    \end{subfigure}
    \begin{subfigure}[t]{0.32\linewidth}
        \includegraphics[width=.95\linewidth]{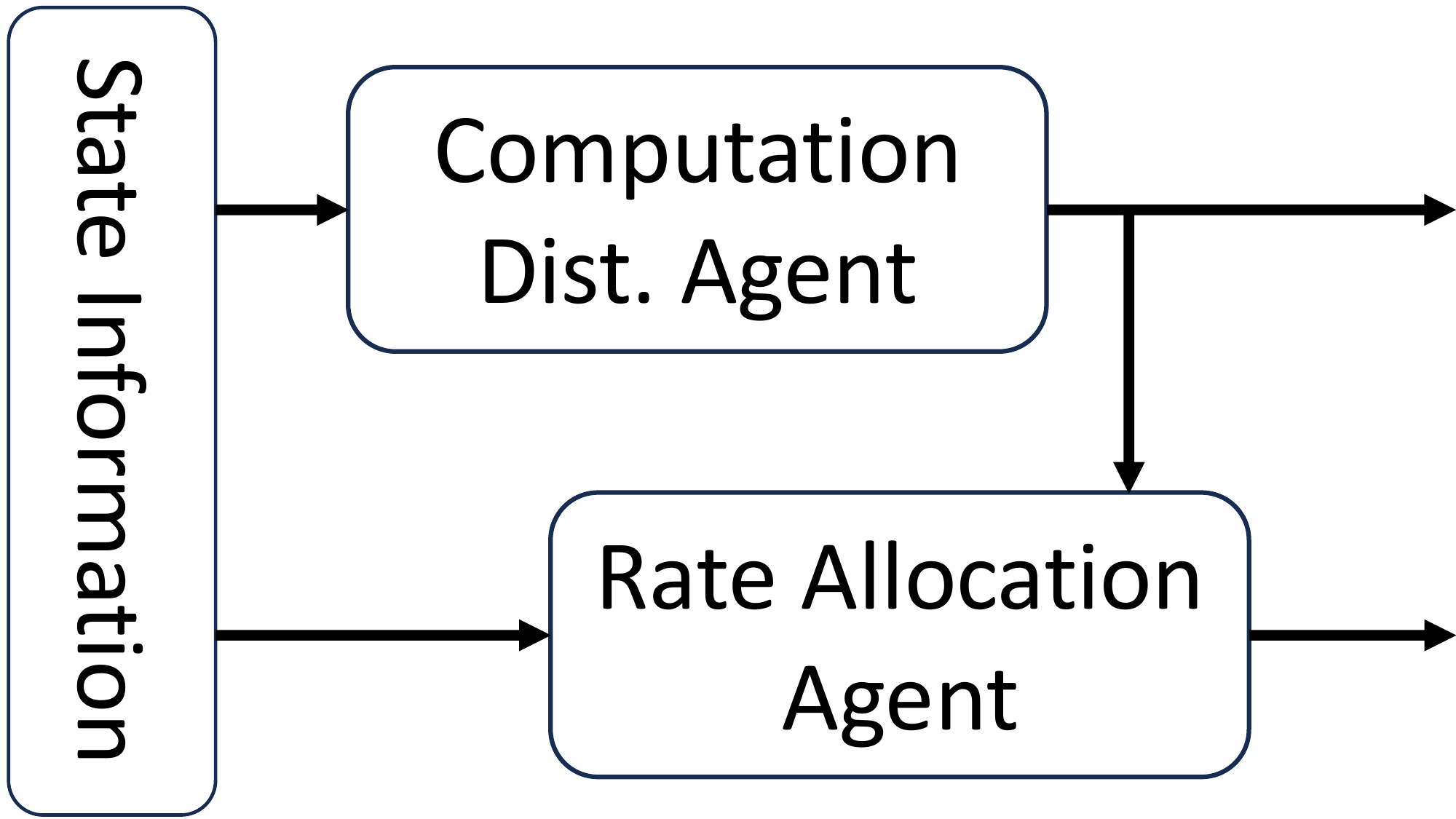}
        \caption{C1R2 Agent Architecture}
        \label{fig:c1r2-arch}    
    \end{subfigure}
    \caption{(a) MTRC makes a joint decision on rate adaptation and computation distribution action. (b) and (c) R1C2 and C1R2 employ neural network cascades to capture the interdependence between the rate adaptation and computation distribution actions.}
    \label{fig:arch}
\end{figure*}

\subsection{MTRC Architecture}

The MTRC agent, as shown in Fig. \ref{fig:deep-vr-arch}, is composed of an actor network $\omega$ and a critic network $\omega_v$.
The actor network outputs the probabilities of both rate allocation action $\pi_{\omega}^{e}$ and computation distribution action $\pi_{\omega}^{\phi}$ for all users.
The actor network also outputs an auxiliary vector that estimates the state value for each of the users separately.
The critic network outputs the estimated state value for each of the users separately.
This architecture reduces the interference between policy and value loss, while distilling features from the value function into the policy network \cite{cobbe-2020-ppg}.
Both actor and critic networks are composed of a convolutional layer and a dense layer.
We employ a three-phase training procedure, consisting of a policy training phase, a coefficient optimization phase, and an auxiliary training phase~\cite{cobbe-2020-ppg, Deheng-2020-Mastering, Tianchi-2023-Buffer}.
\textbf{Policy Optimization:}
In the policy training phase, we update the actor and critic networks.
The policy network is trained by dual-clip proximal policy optimization (PPO) \cite{schulman-2017-proximal}:
\begin{equation}
\small
\label{eq:d-clip-loss}
    \begin{aligned}
        \mathcal{L}^{DClip} = \hat{\mathbb{E}}
            \Big[ \indicator(\hat{A}^m < 0)\max(\mathcal{L}^{PPO}, c\hat{A}^m) + 
            \indicator(\hat{A}^m \geq 0)\mathcal{L}^{PPO} \Big].
    \end{aligned}
\end{equation}
Here, $\indicator(.)$ is a binary indicator function and $\hat{A}^m = r^m + \gamma V_{\omega_v}(s^{m+1}) - V_{\omega_v}(s^m)$ is the advantage function calculated based on the current state value estimate and the discount factor $\gamma = 0.99$, and $\mathcal{L}^{PPO}$ is the surrogate vanilla PPO loss:
\begin{equation}
\small
\label{eq:ppo-loss}
    \begin{aligned}
        \mathcal{L}^{PPO} = \mathcal{L}^{Clip}(\pi_{\omega}, \hat{A}^m) + \kappa H_{\omega}(s^m) + \mathcal{L}^{Value},
    \end{aligned}
\end{equation}
where $H_{\omega}(s^m)$ is the entropy of all the policies and $\kappa$ is the entropy weight.
The entropy loss and its associated weight balance the trade-off between exploration and exploitation in the learning process.
$\mathcal{L}^{Value}$ is the value network loss:
\begin{equation}
\small
\label{eq:value-loss}
    \begin{aligned}
        \mathcal{L}^{Value} &= \mathbb{E}_m\left[ 
            \frac{1}{2}
            (V_{\omega_v}(s^m) - V_{targ}(s^m))^2 
        \right],
    \end{aligned}
\end{equation}
and $\mathcal{L}^{Clip}$ is the single-clip policy loss defined as: 
\begin{equation}
\small
\label{eq:clip-loss}
    \begin{aligned}
        \mathcal{L}^{Clip} = \min \Big[ &\rho(\pi_{\omega}, \pi_{\omega_{old}}) \odot \hat{A}^m, \\ 
 & clip(\rho(\pi_{\omega}, \pi_{\omega_{old}}), 1 \pm \varepsilon) \odot \hat{A}^m \Big].
    \end{aligned}
\end{equation}
Here $\odot$ is the element-wise multiplication, and $\rho(\pi_{\omega}, \pi_{\omega_{old}}) \odot \hat{A}^m$ measures the gained/lost reward as the policy for each user changes, which determines how a change in one user's policy affects other users reward, (\ie perceived QoE).
The change in users' policies w.r.t. their old policy is measured via 
$\rho(\pi_{\omega}, \pi_{\omega_{old}}) = \left[\rho_1, \rho_2, \dotsb, \rho_n \right]$, which is a vector of size $N$.
Each element of this vector $\rho_i$ measures the changes for one user in the environment (\ie the joint probability ratio of the new policy to the old policy for both computation distribution and rate allocation action):
\begin{equation}
\label{eq:rho-i}
\small
    \begin{aligned}
        \rho_i =
        \frac{
            \pi_{\omega}^{\phi} (\phi^m_i|s^m)
        }{
            \pi_{\omega_{old}}^{\phi} (\phi^m_i|s^m)
        } 
        \frac{
            \pi_{\omega}^{e} (e^m_i|s^m)
        }{
            \pi_{\omega_{old}}^{e} (e^m_i|s^m)
        }.
    \end{aligned}
\end{equation}

\textbf{Auxiliary Training Phase:}
In the auxiliary phase, we further optimize the actor and critic networks according to a joint loss function $\mathcal{L}^{Joint}$ on all the experiences.
The joint loss function $\mathcal{L}^{Joint}$ is composed of a behavioral cloning loss and an auxiliary value loss:
\begin{equation}
\small
\label{eq:joint-loss}
    \begin{aligned}
        \mathcal{L}^{Jonit} = 
        \mathbb{E}_m
        \Bigg[ 
            & \mathbf{KL} \left(\pi_{\omega_{old}}^{\phi} (s^m), \pi_{\omega}^{\phi} (s^m)\right)  \\ 
            & + \mathbf{KL} \left(\pi_{\omega_{old}}^{e} (s^m), \pi_{\omega}^{e} (s^m)\right) 
        \Bigg] 
        + \mathcal{L}^{Aux}.
    \end{aligned}
\end{equation}
Here, $\mathbf{KL}(.,.)$ is the behavioral cloning loss, representing the KL-divergence between the original policy and the updated policy.
The $\mathcal{L}^{Aux}$, in Eq. \eqref{eq:joint-loss}, updates the auxiliary value by minimizing the mean squared loss function:
\begin{equation}
\small
\label{eq:aux-loss}
    \begin{aligned}
        \mathcal{L}^{Aux} &= \mathbb{E}_m\left[ 
            \frac{1}{2}
            (V_{\omega}(s^m) - V_{targ}(s^m))^2 
        \right].
    \end{aligned}
\end{equation}


\begin{algorithm}[t]
\caption{MTRC Training Process}
\label{alg:training-alg}

\begin{algorithmic}[1]
    \STATE Initialize $\mathcal{B} \leftarrow \varnothing$, $\mu_0 = 0.1$, $\mu_1 = 0.1$, $\lambda = 0.01$
    \FOR{episode $=1,2,...$}
    \FOR{$m=1,2,...$}
        \STATE $\phi^m \sim \pi_{\omega}^{\phi} (\phi^m|s^m)$ 
        \COMMENT{Sample actions}
        \STATE $e^m \sim \pi_{\omega}^{e} (e^m|s^m)$ 
        \STATE $s^{m+1}$, $r^{m+1}$  = ACT($\phi^m$, $e^m$) 
        \COMMENT{Apply actions}
        \STATE $\mathcal{B} \leftarrow \mathcal{B} \cup \{s^m, (\phi^m, e^m), r^{m+1}, s^{m+1}, V_{targ}(s^m)\}$ 
    \ENDFOR
    \STATE $\mu_0 \leftarrow \mu_0 + \lambda \nabla_{\mu_0}\mathcal{L}^{\mu_0}$ 
    \COMMENT{Update QoE coefficients}
    \STATE $\mu_1 \leftarrow \mu_1 + \lambda \nabla_{\mu_1}\mathcal{L}^{\mu_1}$ 
    \IF{episode $ \%$ $N_{update} = 0 $}
        \FOR{$i = 1,2,..., N_{Policy}$} 
        \COMMENT{Policy training phase}
            \STATE Optimize $\mathcal{L}^{DClip}$ (\ie Eq. \eqref{eq:d-clip-loss}) w.r.t. $\omega, \omega_{v}$
        \ENDFOR
        \FOR{$i = 1,2,..., N_{aux}$} \COMMENT{Auxilary training phase}
            \STATE Optimize $\mathcal{L}^{Value}$ (\ie Eq. \eqref{eq:value-loss}) w.r.t. $\omega_v$
            \STATE Optimize $\mathcal{L}^{Joint}$ (\ie Eq. \eqref{eq:joint-loss}) w.r.t. $\omega$
        \ENDFOR
        \STATE $\mathcal{B} \leftarrow \varnothing$ \COMMENT{Clear replay buffer}
        \STATE $\omega_{old} \leftarrow \omega$, $\omega_{v_{old}} \leftarrow \omega_v$ \COMMENT{Update target network}
    \ENDIF
    \ENDFOR
\end{algorithmic}
\end{algorithm}

\textbf{Training Algorithm:}
Algorithm~\ref{alg:training-alg} presents the training process of the MTRC agent.
The training continues for multiple iterations until convergence.
Each iteration consists of four phases.
In the first phase, we perform the current policy $\pi_{\omega}$ on a randomized environment to collect new experiences (\ie rollout process).
In the rollout process (lines 3-8), we first sample rate adaptation and computation distribution actions, and then apply the chosen actions on the edge-assisted VR environment. 
Then, we store the resulted transition in a replay buffer for training purposes.
In the second phase (\ie coefficient optimization phase), we update the rebuffering time and quality variation coefficients 
using the loss functions defined for the coefficients in Eq. \eqref{eq:qoe-lagrange} (lines 9-10).
In the third phase (\ie policy training phase), we update both the actor and critic networks.
We use a random batch from the replay buffer to compute the dual-clip PPO loss $\mathcal{L}^{DClip}$, and update the networks (lines 11-14).
Finally, in the last phase (\ie auxiliary training phase), we further update the actor and critic networks by optimizing the behavioral cloning and value losses using all the replay buffer data (lines 15-18).



\subsection{Multitask Neural Network Cascades}
To extend the MTRC framework, we propose multitask neural network cascades for rate adaptation and computation distribution~\cite{Dai-2016-Instance}.
Our neural network cascades (\ie R1C2 and C1R2) have two stages, each of which addresses one sub-task (\ie rate adaptation and computation distribution).
The two stages share the state information. However, unlike prior multitask RL frameworks (e.g.,~\cite{Deheng-2020-Mastering}), in our method, a later stage depends on the output of an earlier stage, as shown in Fig. \ref{fig:arch}. This forms a causal cascade, which helps to capture the interdependence between the decision variables in the optimization problem formulated in Eq. \eqref{eq:qoe-maximization}.
However, since the direction of the causal dependence is not trivial, we present two variants \bb{of} multitask neural network cascades (\ie R1C2 and C1R2).

As shown in Fig. \ref{fig:r1c2-arch}, the R1C2 architecture first decides on the rate allocated to each user in the environment. 
Then, the rate allocation actions along with the state information are fed to the computation distribution agent. 
R1C2 captures the dependence of the computation distribution action on the rate allocation action. In contrast,
C1R2 (see Fig. \ref{fig:c1r2-arch}) first decides on the computation distribution (\ie where each GoP needs to be decoded/rendered). 
Then, the computation distribution actions combined with the state information are fed to the rate adaptation agent to decide how many enhancement layers should be processed for each user. As such, the C1R2 architecture captures the dependence of
the rate adaptation action on the computation distribution action.



\textbf{Policy Optimization:}
Similar to the MTRC agent, we employ a three-phase training process for the R1C2 and C1R2 agents, but we split Eq. \eqref{eq:rho-i} into two parts in order to separately measure the rate adaptation and the computation distribution policy changes using $\rho_i^{e}$ and $\rho_i^\phi$, respectively. Thus, we have: 
\begin{equation}
\label{eq:rho-i-e}
\small
    \begin{aligned}
        \rho_i^{e} =
        \frac{
            \pi_{\omega}^{e} (e^m_i|s^m)
        }{
            \pi_{\omega_{old}}^{e} (e^m_i|s^m)
        }, \quad
        \rho_i^\phi =
        \frac{
            \pi_{\omega}^{\phi} (\phi^m_i|s^m)
        }{
            \pi_{\omega_{old}}^{\phi} (\phi^m_i|s^m)
        } 
        .
    \end{aligned}
\end{equation}
\noindent 
Similarly, in the auxiliary training phase, Eq. \eqref{eq:joint-loss} is split into two parts. 
The rate adaptation agent only uses the behavioral cloning loss for rate adaptation action and auxiliary value loss:
\begin{equation}
\small
\label{eq:joint-loss-e}
    \begin{aligned}
        \mathcal{L}^{Jonit}_e = 
        \mathbb{E}_m
        \Big[ 
            & \mathbf{KL} \left(\pi_{\omega_{old}}^{e} (s^m), \pi_{\omega}^{e} (s^m)\right) 
        \Big] 
        + \mathcal{L}^{Aux},
    \end{aligned}
\end{equation}
and the computation distribution agent uses behavioral cloning loss for computation distribution action along with auxiliary value loss to further improve the value estimation:
\begin{equation}
\small
\label{eq:joint-loss-phi}
    \begin{aligned}
        \mathcal{L}^{Jonit}_\phi = 
        \mathbb{E}_m
        \Big[
            & \mathbf{KL} \left(\pi_{\omega_{old}}^{\phi} (s^m), \pi_{\omega}^{\phi} (s^m)\right)  
        \Big] 
        + \mathcal{L}^{Aux}.
    \end{aligned}
\end{equation}
These modifications result in some changes in Algorithm \ref{alg:training-alg}. In line 13, the dual-clip PPO loss $\mathcal{L}^{DClip}$ is replaced with two separate loss functions (\ie policy loss for the computation distribution agent $\mathcal{L}^{DClip}_\phi$, and policy loss for rate adaptation agent $\mathcal{L}^{DClip}_e$). Moreover, $\mathcal{L}^{Jonit}$ in line 17 is replaced with Eq. \eqref{eq:joint-loss-e} and \eqref{eq:joint-loss-phi}. 
In the next section, we evaluate the performance of the proposed neural network architectures.


\section{Evaluation}\label{sec:evaluation}


In this section, we evaluate our proposed framework through an extensive simulation 
against two learning-based methods, \ie  Pensieve \cite{Mao-2017-Neural} and COREL \cite{Hu-2023-COREL}, and a buffer-based adaptive-bitrate (ABR) algorithm \cite{Huang-2014-BBA}.
Both Pensieve and COREL only adjusts the video rate based on user state, while our problem is a joint computation distribution and rate adaptation algorithm. 
Thus, we employ two variants of these methods. The first variants of these methods, namely ECU-COREL, ECU-Pensieve and ECU-BBA,
perform all the computations (\ie decoding and rendering) on the ECU.
The second variants, named Headset-COREL, Headset-Pensieve, and Headset-BBA, however perform all the computations on the users' headsets.
Moreover, we present two rate adaptation algorithms, ECU-R and Headset-R, based on our multitask framework.
ECU-R and Headset-R use a neural network with the same architecture as shown in Fig. \ref{fig:deep-vr-arch} for rate adaptation, except that the computation distribution is not decided by the neural network and all computations are performed on the ECU or headsets, respectively.



\begin{table}[h]
    \centering
    \resizebox{.55\linewidth}{!}{
    \begin{tabular}{lc}
        \toprule
        Definition/Explanation & Parameter \& Value \\
        \midrule
        Number of VR users & $N=6$ \\
        Rendering/compression factors & $\alpha, \beta = 2.1, 0.6$ \\
        User prefered rebuffering time & $\mathcal{H}_0 = 2$ Seconds\\
        Max allowed quality variation & $1.09 \leq \mathcal{H}_1 \leq 2.99$ dB \\
        ECU decoding/rendering speed & $Z^{dec.}_{ECU}, Z^{rend.}_{ECU} = 7.5, 20\,Gbps$ \\
        User decoding/rendering speed & $Z^{dec.}_{n}, Z^{rend.}_{n}=0.2, 9.4 \, Gbps$ \\
        \midrule
        Number of training iterations & $N_{policy}=  80$, $N_{aux}= 6$ \\
        Initial RT \& QV coefficient & $\mu_0=0.1$, $\mu_1=0.1$ \\
        Policy update frequency & $N_{update}=4$ \\
        Entropy weight & $\kappa=0.01$ \\
        \bottomrule
    \end{tabular}
    }
    \caption{Simulation and training parameter values.}
    \label{tab:parameter-values}
\end{table}

\textbf{Datasets and Training:} In our simulations, we employ a full UHD $360^\circ$ video dataset \cite{Chakareski-2021-Full}.
This dataset includes 15 videos with various spatio-temporal characteristics.
Each video is represented using the multi-layer $360^\circ$ model presented in Section \ref{sec:system}, and video frames are partitioned into an $8\times8$ grid.
Bitrate information for seven layers, each offering progressively higher levels of immersion fidelity for each tile, is provided.
Additionally, head movement data for multiple users is included, enabling us to determine the viewport location for each user and simulate the multi-user VR arena.
Moreover, we use a dataset of mmWave network throughput traces~\cite{Narayanan-2021-Variegated}, which were collected in two different cities in the U.S. and from commercial \bb{5G} operators (T-Mobile and Verizon). 
We employ these datasets to train all the agents mentioned above for $5,000$ episodes. The $360^\circ$ videos and network traces are randomly chosen in each episode.
To ensure a fair comparison, we use the same configuration for all the baselines as presented in Table \ref{tab:parameter-values}.

\begin{figure}[t]
\centering
    \begin{subfigure}{\linewidth}
         \centering
         \includegraphics[width=.6\linewidth]{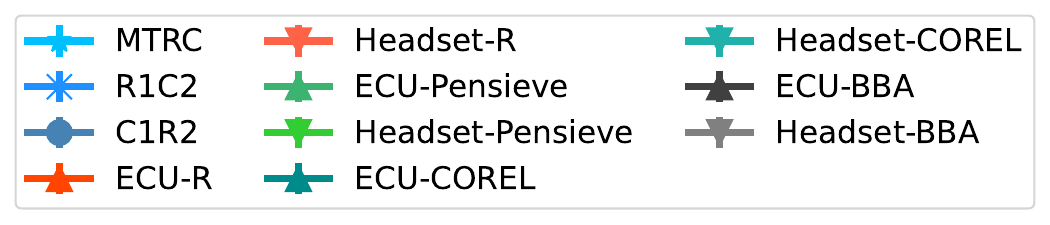}
         \vspace{-.035in}
     \end{subfigure}
    \begin{subfigure}[t]{0.48\linewidth}
         \centering
         \includegraphics[width=\linewidth]{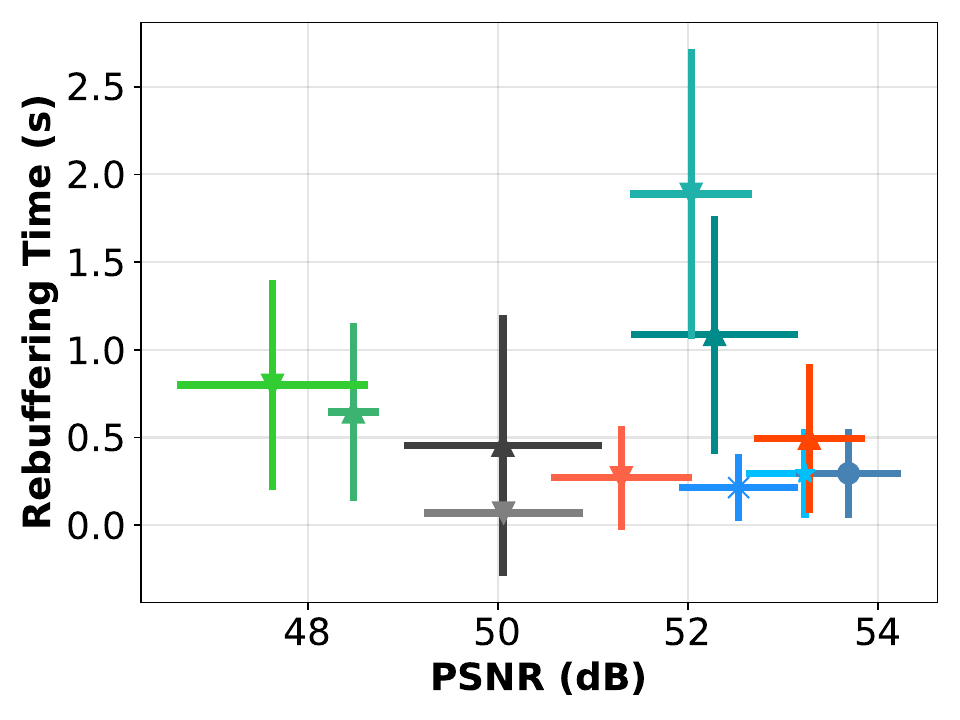}
         \caption{RT vs. PSNR}
     \end{subfigure}
     \begin{subfigure}[t]{0.48\linewidth}
         \centering
         \includegraphics[width=\linewidth]{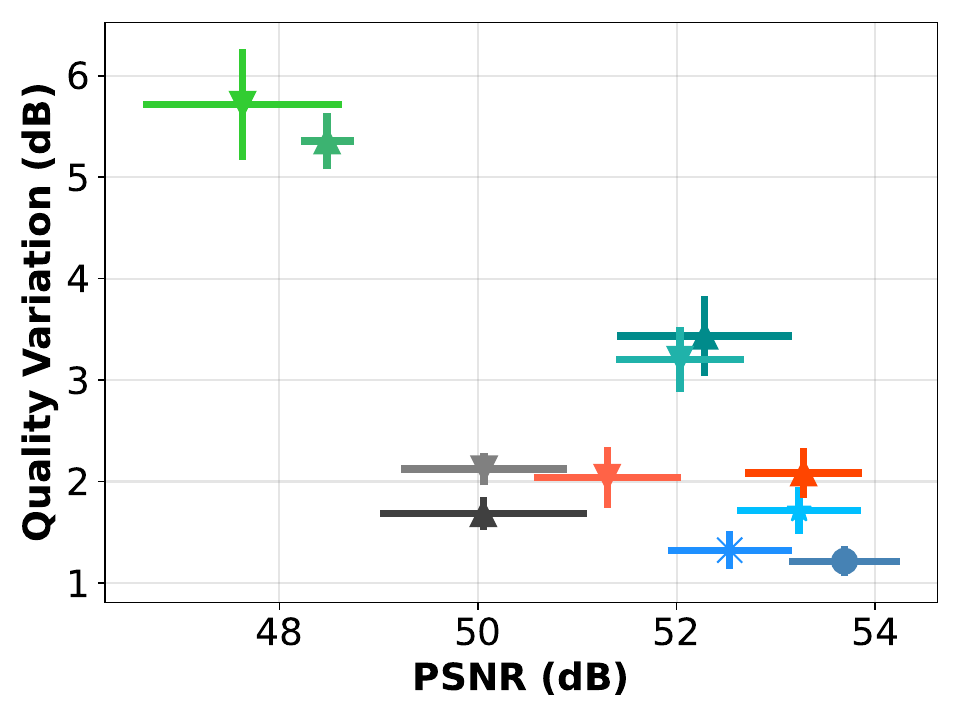}
         \caption{QV vs. PSNR}
     \end{subfigure}
\caption{Performance trade-offs between rebuffering time (RT), quality variation (QV), and PSNR during testing stage.}
\label{fig:testing-performance}
\end{figure}

\textbf{Deployment Performance:}
Fig. \ref{fig:testing-performance} demonstrates the performance trade-offs between rebuffering time, quality variation, and PSNR, which is collected from 300 episodes of testing.
Each point demonstrates the average rebuffering time (or quality variation) and PSNR experienced by the users. The vertical and horizontal bars represent the standard deviation of the rebuffering time (or quality variation) and PSNR, respectively.
A small rebuffering time (or quality variation) and high PSNR with small variation is desirable, which is represented by a point in the lower right corner of these plots. From the results, we observe that MTRC, R1C2, and C1R2 agents provide the best trade-off, with the C1R2 agent outperforming all other baselines. 
Specifically, the C1R2 agent, compared to ECU-Pensieve and Headset-Pensieve, achieves $5.21-6.06$ dB PSNR gains, $2.18-2.70$x rebuffering time reduction, and $4.14-4.50$ dB quality variation reduction.
Moreover, C1R2 agents, compared to ECU-COREL and Headset-COREL, achives $1.41-1.65$ db in PSNR gain, $3.67-6.39$x rebuffering time reduction, and $1.99-2.22$ db quality variation reduction.
These performance gains are due to two main reasons.
First, our proposed agents can dynamically choose where to prepare a GoP and distribute the computations on the ECU and headsets. This, in turn, better balances the communication and computation trade-off compared with the Pensieve COREL variations.
Second, Pensieve and COREL variations have no information about the preparation process, thus unable to capture the relation between processing times (\ie decoding and rendering times) and their impacts on QoE.

Furthermore, Table \ref{tab:video-stats} reports the average and standard deviation of PSNR, rebuffering time, and quality variations for groups of users that play a specific video. 
We see that C1R2 provides the best PSNR and quality variation for various spatio-temporal characteristics.
This is due to the fact that the rate adaptation agent has access to the computation distribution action, which means that it knows how the computations are distributed across the ECU and VR devices.
Thus, it allocates rates to users based on the available computational resources.
However, R1C2, although not significantly different, provides the best rebuffering time for various spatio-temporal characteristics.
This demonstrates that the computation distribution agent in R1C2 uses rate allocation actions to distribute computations such that the rebuffering time is minimized.


\begin{figure}
    \centering
\includegraphics[width=.6\linewidth]{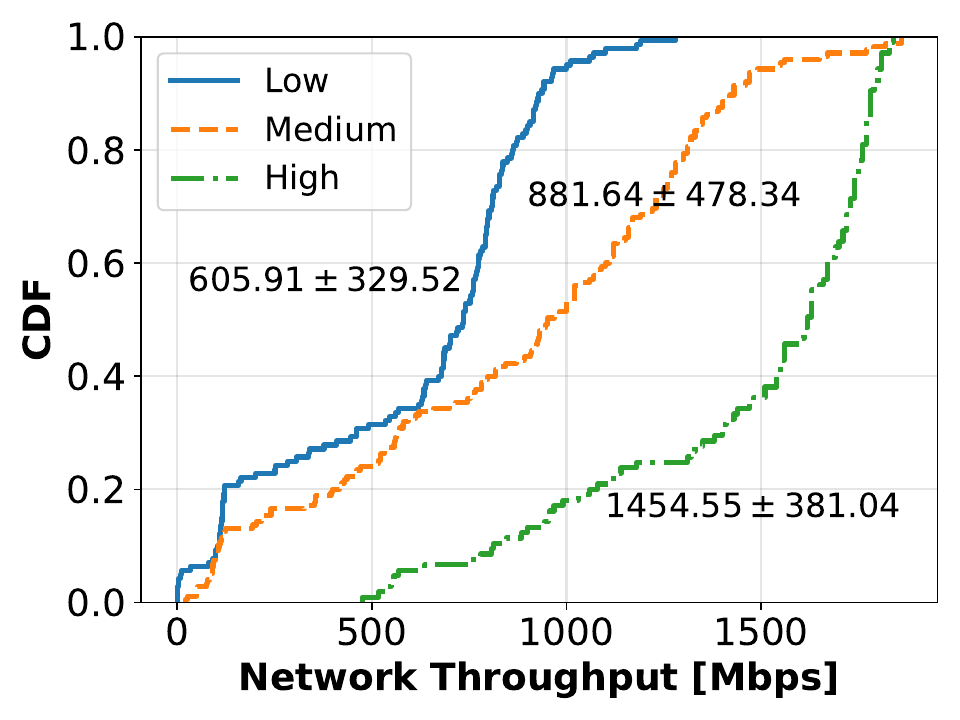}
    \vspace{-.1in}
    \caption{Network throughput cumulative distribution function (CDF).}
    \label{fig:data-rate-pdf}
\end{figure}

\begin{table*}[h]
    \centering
    \resizebox{.95\linewidth}{!}{
    \begin{tabular}{l|c|c|c|c|c|c|c|c|c}
         \textbf{} & \multicolumn{3}{c|}{\textbf{MTRC}} & \multicolumn{3}{c|}{\textbf{C1R2}} & \multicolumn{3}{c}{\textbf{R1C2}} \\
        \hline
        \textbf{Video Name} & \textbf{PSNR (dB)} & \textbf{RT (Sec)} & \textbf{QV (dB)} & \textbf{PSNR (dB)} & \textbf{RT (Sec)} & \textbf{QV (dB)} & \textbf{PSNR (dB)} & \textbf{RT (Sec)} & \textbf{QV (dB)} \\
        \hline
        Academic & $52.05 \pm 1.10$ & $0.13 \pm 0.32$ & $1.58 \pm 0.41$ & $\bgreen{52.40 \pm 1.04}$ & $0.10 \pm 0.18$ & $\bblue{1.08 \pm 0.11}$ & $51.42 \pm 1.17$ & $\bred{0.09 \pm 0.16}$ & $1.12 \pm 0.26$ \\
        Basketball & $53.72 \pm 1.34$ & $0.34 \pm 0.60$ & $2.29 \pm 0.44$ & $\bgreen{54.25 \pm 1.11}$ & $0.27 \pm 0.47$ & $\bblue{1.93 \pm 0.16}$ & $52.92 \pm 1.72$ & $\bred{0.24 \pm 0.31}$ & $2.09 \pm 0.41$ \\
        Bridge & $55.00 \pm 0.85$ & $0.57 \pm 1.17$ & $1.27 \pm 0.36$ & $\bgreen{55.22 \pm 1.27}$ & $0.57 \pm 1.00$ & $\bblue{0.80 \pm 0.14}$ & $54.05 \pm 1.39$ & $\bred{0.36 \pm 0.71}$ & $0.88 \pm 0.23$ \\
        Gate Night & $55.39 \pm 1.09$ & $\bred{0.25 \pm 0.62}$ & $1.33 \pm 0.35$ & $\bgreen{55.63 \pm 1.29}$ & $0.39 \pm 1.04$ & $\bblue{0.95 \pm 0.16}$ & $54.62 \pm 1.35$ & $0.28 \pm 0.58$ & $1.03 \pm 0.36$ \\
        Runner & $52.56 \pm 0.78$ & $0.25 \pm 0.57$ & $1.81 \pm 0.40$ & $\bgreen{52.80 \pm 0.83}$ & $0.25 \pm 0.57$ & $\bblue{1.37 \pm 0.14}$ & $52.02 \pm 1.01$ & $\bred{0.22 \pm 0.51}$ & $1.55 \pm 0.29$ \\
        Siyuan & $52.96 \pm 0.98$ & $0.10 \pm 0.26$ & $1.88 \pm 0.40$ & $\bgreen{53.21 \pm 1.00}$ & $\bred{0.08 \pm 0.17}$ & $\bblue{1.50 \pm 0.13}$ & $52.35 \pm 1.08$ & $0.08 \pm 0.25$ & $1.58 \pm 0.22$ \\
        South Gate & $53.22 \pm 1.47$ & $0.26 \pm 0.65$ & $1.99 \pm 0.48$ & $\bgreen{53.86 \pm 1.49}$ & $0.23 \pm 0.54$ & $\bblue{1.41 \pm 0.16}$ & $52.23 \pm 1.96$ & $\bred{0.15 \pm 0.39}$ & $1.54 \pm 0.32$ \\
        Studyroom & $54.23 \pm 1.06$ & $0.13 \pm 0.33$ & $1.36 \pm 0.38$ & $\bgreen{54.57 \pm 1.09}$ & $0.14 \pm 0.40$ & $\bblue{0.90 \pm 0.11}$ & $53.51 \pm 1.11$ & $\bred{0.09 \pm 0.21}$ & $0.97 \pm 0.15$ \\
        Sward & $50.28 \pm 1.97$ & $0.68 \pm 1.02$ & $1.83 \pm 0.62$ & $\bgreen{51.32 \pm 1.29}$ & $0.64 \pm 0.87$ & $\bblue{1.04 \pm 0.17}$ & $49.80 \pm 2.02$ & $\bred{0.45 \pm 0.58}$ & $1.11 \pm 0.34$ \\
    \end{tabular}
    }
    \caption{\centering QoE perceived by groups of users who play a specific video. The bold black, red, and blue values, respectively, show the best PSNR, rebuffering time, and quality variation achieved for each video. }
    \label{tab:video-stats}
\end{table*}

\textbf{Effect of Wireless Network Condition:}
We also demonstrate the impact of the wireless network condition by analyzing the QoE perceived by three groups of users under various network conditions, including low, medium, and high throughput scenarios. Fig. \ref{fig:data-rate-pdf} shows the  statistical characteristics (\ie cumulative distribution function (CDF), average, and standard deviation) of these three groups.
The results in Tables~\ref{tab:psnr-table}, \ref{tab:rebuffering-time-table}, and \ref{tab:quality-var-table} demonstrate the average and standard deviation of the PSNR, rebuffering time, and quality variation for these three groups of users.
\emph{(1) PSNR Performance:}
In Table \ref{tab:psnr-table}, we observe that C1R2 provides the best results in all network conditions.
We also note that as the network condition improves (\ie higher throughput), the PSNR increases.
Furthermore, C1R2 provides a smaller variation in PSNR as the network condition changes ($53.65$ dB for low-throughput users compared to $53.70$ dB for high-throughput users).
\emph{(2) Rebuffering Time Performance:}
In Table \ref{tab:rebuffering-time-table}, we can see that the rebuffering time improves as the network condition improves.
Moreover, R1C2 provides the best rebuffering time across all users. 
\emph{(3) Quality Variation Performance:}
As shown in Table \ref{tab:quality-var-table}, although C1R2 provides the best quality variation, the quality variation does not change significantly as the network condition changes.

\begin{table}[htbp]
    \centering
    \resizebox{.5\linewidth}{!}{
    \begin{tabular}{|l|c|c|c|}
        \hline
        \multicolumn{4}{|c|}{\textbf{PSNR [dB]}} \\
        \hline
        \textbf{\diagbox[width=9em]{Baseline}{Throughput}} & \textbf{Low} & \textbf{Medium} & \textbf{High} \\
        \hline
        MTRC & $53.15 \pm 1.81$ & $53.39 \pm 1.86$ & $53.57 \pm 1.75$ \\
        C1R2 & $\mathbf{53.65 \pm 1.85}$ & $\mathbf{53.67 \pm 1.70}$ & $53.70 \pm 1.75$ \\
        R1C2 & $52.32 \pm 2.08$ & $52.47 \pm 1.96$ & $52.91 \pm 1.79$ \\
        ECU-R & $53.17 \pm 1.54$ & $53.27 \pm 1.57$ & $\mathbf{53.71 \pm 1.55}$ \\
        Headset-R & $51.21 \pm 2.31$ & $51.21 \pm 2.21$ & $51.52 \pm 1.95$ \\
        ECU-Pensieve & $48.43 \pm 1.60$ & $48.49 \pm 1.75$ & $48.36 \pm 1.56$ \\
        Headset-Pensieve & $47.72 \pm 2.43$ & $47.45 \pm 2.48$ & $47.61 \pm 2.56$ \\
        {ECU-COREL} & {$51.93 \pm 2.02$} & {$52.56 \pm 1.98$} & {$53.29 \pm 1.90$} \\
{Headset-COREL} & {$51.92 \pm 1.74$} & {$52.05 \pm 1.73$} & {$52.42 \pm 1.75$} \\
        {ECU-BBA} & {$48.57 \pm 3.96$} & {$50.06 \pm 2.20$} & {$50.45 \pm 2.15$} \\
        {Headset-BBA} & {$49.69 \pm 2.24$} & {$50.03 \pm 2.04$} & {$50.41 \pm 1.92$} \\
        \hline
    \end{tabular}
    }
    \caption{Effect of network condition on PSNR. The bold values denote the baseline that provides the best PSNR.}
    \label{tab:psnr-table}
\end{table}

\begin{table}[t]
    \centering
    \resizebox{.5\linewidth}{!}{
    \begin{tabular}{|l|c|c|c|}
        \hline
        \multicolumn{4}{|c|}{\textbf{Rebuffering Time [Seconds]}} \\
        \hline
        \textbf{\diagbox[width=9em]{Baseline}{Throughput}} & \textbf{Low} & \textbf{Medium} & \textbf{High} \\
        \hline
        MTRC & $0.49 \pm 0.85$ & $0.30 \pm 0.59$ & $0.06 \pm 0.04$ \\
        C1R2 & $0.58 \pm 1.16$ & $0.27 \pm 0.52$ & $0.07 \pm 0.07$ \\
        R1C2 & $\mathbf{0.38 \pm 0.75}$ & $\mathbf{0.21 \pm 0.40}$ & $\mathbf{0.06 \pm 0.04}$ \\
        ECU-R & $0.92 \pm 1.23$ & $0.51 \pm 0.89$ & $0.06 \pm 0.05$ \\
        Headset-R & $0.59 \pm 1.19$ & $0.32 \pm 0.82$ & $0.09 \pm 0.11$ \\
        ECU-Pensieve & $1.34 \pm 1.83$ & $0.55 \pm 1.12$ & $0.06 \pm 0.04$ \\
        Headset-Pensieve & $1.28 \pm 1.82$ & $0.80 \pm 1.44$ & $0.40 \pm 0.77$ \\
        {ECU-COREL} & {$2.05 \pm 1.64$} & {$0.86 \pm 1.35$} & {$0.06 \pm 0.04$} \\
        {Headset-COREL} & {$2.75 \pm 2.81$} & {$1.67 \pm 1.85$} & {$2.19 \pm 2.51$} \\
        {ECU-BBA} & {$2.54 \pm 4.56$} & {$0.08 \pm 0.06$} & {$0.07 \pm 0.05$} \\
        {Headset-BBA} & {$0.07 \pm 0.05$} & {$0.08 \pm 0.05$} & {$0.06 \pm 0.04$} \\
        \hline
    \end{tabular}
    }
        \caption{Effect of network condition on rebuffering time. The bold values denote the baseline that provides the best rebuffering time.}
    \label{tab:rebuffering-time-table}
\end{table}

\begin{table}[t]
    \centering
    \resizebox{.5\linewidth}{!}{
    \begin{tabular}{|l|c|c|c|}
        \hline
        \multicolumn{4}{|c|}{\textbf{Quality Variation [dB]}} \\
    \hline
    \textbf{\diagbox[width=9em]{Baseline}{Throughput}} & \textbf{Low} & \textbf{Medium} & \textbf{High} \\
        \hline
        MTRC & $1.76 \pm 0.54$ & $1.70 \pm 0.51$ & $1.60 \pm 0.48$ \\
        C1R2 & $\mathbf{1.23 \pm 0.38}$ & $\mathbf{1.20 \pm 0.35}$ & $\mathbf{1.21 \pm 0.38}$ \\
        R1C2 & $1.41 \pm 0.58$ & $1.31 \pm 0.42$ & $1.27 \pm 0.43$ \\
        ECU-R & $2.11 \pm 0.58$ & $2.07 \pm 0.63$ & $1.90 \pm 0.56$ \\
        Headset-R & $2.06 \pm 1.13$ & $2.05 \pm 1.12$ & $2.00 \pm 1.15$ \\
        ECU-Pensieve & $5.31 \pm 1.07$ & $5.35 \pm 1.17$ & $5.43 \pm 1.04$ \\
        Headset-Pensieve & $5.67 \pm 1.35$ & $5.73 \pm 1.42$ & $5.76 \pm 1.44$ \\
        {ECU-COREL} & {$3.73 \pm 0.95$} & {$3.26 \pm 0.84$} & {$2.76 \pm 0.81$} \\
        {Headset-COREL} & {$3.11 \pm 0.76$} & {$3.19 \pm 0.78$} & {$3.16 \pm 0.82$} \\
        {ECU-BBA} & {$1.89 \pm 0.52$} & {$1.58 \pm 0.26$} & {$1.28 \pm 0.29$} \\
        {Headset-BBA} & {$2.16 \pm 0.44$} & {$2.12 \pm 0.39$} & {$2.23 \pm 0.43$} \\
        \hline
    \end{tabular}
    }
    \caption{Effect of network condition on video quality variation. The bold values denote the baseline that provides the best quality variation.}
    \label{tab:quality-var-table}
\end{table}

\textbf{Video Quality Assessment}:
Performance of streaming applications is evaluated using subjective and objective tests. 
Subjective tests are used to evaluate the user's perceptual experience as they aim to capture the actual streaming experience of the user, which depends on a variety of factors.
For example, subjective tests can provide feedback on aspects such as how engaging and realistic the VR experience felt.
Moreover, subjective tests can reveal the user's preferences (e.g., the preferred quality vs. latency trade-offs, sensitivity to rebuffering or lag in immersive VR environment, and tolerance for reduced resolution).
However, subjective tests are more challenging and costly to conduct, as they require human participants, and they provide qualitative data, which is harder to quantify, analyze, and reproduce due to its nature and the diversity of the participants. Moreover, they lack a direct and analytically quantifiable connection to optimization objectives in streaming systems design, and thus cannot benefit related studies.

To overcome the challenges of performing subjective tests, 
objective metrics, such as the mean squared error (MSE) and PSNR, have been proposed to measure the quality of streaming/encoding algorithms for video applications.
MSE and PSNR are widely used due to various reasons.
MSE and PSNR are easy to calculate, have a clear physical meaning, and are mathematically convenient for optimization \cite{Zhou-2004-SSIM}.
However, there are some aspect of visual quality that are not captured by MSE and PSNR.
Hence, other metrics have been proposed in the literature that integrate other aspects of visual quality \cite{Zhou-2004-SSIM} and subjective measures  \cite{aaron-2015-VMAF} into consideration.

Structural similarity index measure (SSIM) integrates the visual quality of the images by considering image degradations as perceived changes in structural information, luminance, and contrast~\cite{Zhou-2004-SSIM}.
Furthermore, Netflix has integrated more subjective measures into consideration by 
introducing Video Multi-method Assessment Fusion (VMAF) metric~\cite{aaron-2015-VMAF}.
VMAF employs a machine learning model to combine various video features (e.g., spatial quality, temporal quality, and preservation of accurate color representation) into a single quality score. This model is trained on a subjective video quality dataset where human viewers have rated videos for perceptual quality.
Therefore, VMAF is able to integrate various perceptual aspects,
while providing an objective metric to evaluate the performance of streaming/encoding methods.



Within this context, Table \ref{tab:subjective-test} reports the average and standard deviation of SSIM, VMAF, MSE, and PSNR for a subset of users in the environment.
The SSIM column shows a small variation in SSIM across different baselines.
This small variation indicates that all approaches are effective in preserving the structural integrity of the video content, suggesting that while there are differences in overall quality metrics like VMAF, MSE, and PSNR, the core visual structure remains largely consistent across these techniques.

Notably, C1R2 achieves the highest scores in VMAF ($95.60 \pm 1.96$) and MSE ($0.22 \pm 0.14$), indicating its ability to maintain video quality and minimize distortion. 
Although MTRC and R1C2 slightly underperform compared with the C1R2 method, they consistently outperform Pensieve, COREL, and BBA variations. For instance, Pensieve-based approaches, like ECU-Pensieve and Headset-Pensieve, showed lower PSNR and VMAF values, and significantly higher MSE, highlighting their limitations in maintaining consistent video quality.
In contrast, our proposed methods, particularly C1R2, deliver robust and stable performance, demonstrating their effectiveness in providing high-quality video experiences. This comparison underscores the efficiency of our approaches in optimizing visual quality and resilience against distortion, making them more suitable for applications requiring high precision and reliability.


\begin{table}[ht]
\centering
\resizebox{.5\linewidth}{!}{
\begin{tabular}{|l|c|c|c|c|c|}
\hline
\textbf{Baseline} & \textbf{SSIM} & \textbf{VMAF} & \textbf{MSE} & \textbf{PSNR [dB]} \\ 
\hline
MTRC & $0.99 \pm 0.01$ & $95.49 \pm 2.92$ & $0.23 \pm 0.23$ & $\mathbf{56.33 \pm 3.23}$ \\ 
C1R2 & $\mathbf{0.99 \pm 0.00}$ & $\mathbf{95.60 \pm 1.96}$ & $\mathbf{0.22 \pm 0.14}$ & $56.04 \pm 2.73$ \\ 
R1C2                  & $0.99 \pm 0.01$ & $94.78 \pm 3.31$ & $0.30 \pm 0.24$ & $54.95 \pm 3.18$ \\ 
ECU-R                 & $0.99 \pm 0.00$ & $95.49 \pm 2.13$ & $0.23 \pm 0.15$ & $55.86 \pm 2.84$ \\ 
Headset-R             & $0.99 \pm 0.01$ & $93.83 \pm 3.72$ & $0.39 \pm 0.26$ & $53.49 \pm 3.06$ \\ 
ECU-Pensieve          & $0.98 \pm 0.01$ & $90.01 \pm 7.41$ & $0.70 \pm 0.52$ & $51.03 \pm 5.14$ \\ 
Headset-Pensieve      & $0.98 \pm 0.02$ & $89.92 \pm 7.38$ & $0.72 \pm 0.52$ & $50.86 \pm 5.28$ \\ 
ECU-COREL             & $0.99 \pm 0.01$ & $94.72 \pm 4.31$ & $0.29 \pm 0.32$ & $55.62 \pm 3.76$ \\ 
Headset-COREL         & $0.99 \pm 0.01$ & $94.56 \pm 3.57$ & $0.32 \pm 0.26$ & $54.68 \pm 3.39$ \\ 
ECU-BBA & $0.99 \pm 0.01$ & $92.91 \pm 6.55$ & $0.45 \pm 0.51$ & $54.36 \pm 5.39$ \\
Headset-BBA & $0.99 \pm 0.01$ & $92.94 \pm 5.85$ & $0.45 \pm 0.45$ & $53.46 \pm 4.16$ \\       
\hline
\end{tabular}
}
\caption{Perceived QoE in terms of SSIM, VMAF, YSME, and PSNR. The bold values denote the baseline that provides the best performance.}
\label{tab:subjective-test}
\end{table}

\section{Conclusion}\label{sec:conclusion}

In this paper, we considered the problem of multitask rate adaptation and computation distribution in a VR arena for a $360^\circ$ video streaming platform. We present a learning-based multitask agent that decides on the video bitrate allocated to each user and computation distribution (\ie whether each video segment is decoded/rendered on the ECU or on the headset).
The overall objective is to maximize the video quality of users under dynamic and time-varying conditions in terms of video requests, available computational resources, communication bandwidth, and user requirements. 
Using the state-of-the-art DRL algorithm, we developed MTRC that utilizes playback statistics and video information to make a joint rate adaptation and computation distribution decision. 
Furthermore, we leverage neural network cascades to extend our MTRC and introduce the R1C2 and C1R2 methods.
R1C2 and C1R2 further improve the QoE for VR users by accounting for the interdependence between the rate adaptation action and the computation distribution action.
Through numerical simulation using real-world network traces and $360^\circ$ video information, we showed that the proposed methods learn to balance the existing trade-offs in the system and outperform the state-of-the-art rate adaptation algorithms. 
Specifically, the C1R2 agent demonstrated the best performance and achieved $5.21-6.06$ dB PSNR gains, $2.18-2.70$x rebuffering time reduction, and $4.14-4.50$ dB quality variation reduction. 

%

\section{Acknowledgment}
This work was supported in part by NSF grants 1955561, 2212565, 2323189, 2032033, 2106150, and 2346528. The work of Jacob Chakareski was additionally supported in part by NIH award R01EY030470; and by the Panasonic Chair of Sustainability at NJIT.

{\footnotesize
\bibliographystyle{IEEEtranN}
\bibliography{ref-combined}
}

\end{document}